\documentclass[10pt, 3p]{elsarticle-mod}

\usepackage[english]{babel}
\usepackage[utf8]{inputenc}
\usepackage{graphics}
\usepackage{graphicx}
\usepackage{amsmath,amssymb,esint}
\usepackage{lineno}
\usepackage{multirow}
\usepackage{subfig}
\usepackage{natbib}
\usepackage{eucal}
\usepackage{mathtools}
\usepackage{placeins}
\usepackage{enumerate}
\usepackage[usenames,x11names]{xcolor}
\usepackage[colorlinks]{hyperref}
\usepackage{setspace}
\usepackage{MnSymbol}
\usepackage{upgreek}
\usepackage{listings}
\usepackage{amsmath,amssymb}
\usepackage{algorithm}
\usepackage{algpseudocode}
\usepackage{float}
 
\lstdefinestyle{cpp}{
    language=C++,
    basicstyle=\ttfamily\footnotesize,
    columns=fullflexible,
    keepspaces=true,
    showstringspaces=false,
    breaklines=true,
    breakatwhitespace=false,
    frame=single,
    numbers=left,
    numberstyle=\scriptsize,
    numbersep=6pt,
    tabsize=4
}

\biboptions{sort&compress}
\date{}

\begin{document}

\begin{frontmatter}

\title{QMOMKit: A C++ library for quadrature-based moment methods}

\author{Fabian Denner\corref{cor1}}
\ead{fabian.denner@polymtl.ca}
\author{Bruna Campos}
\author{Bruno Blais}
\address{Polytechnique Montréal, Montréal, H3T 1J4, Québec, Canada}
\cortext[cor1]{Corresponding author: }

\begin{abstract}
QMOMKit is an open-source C++ library for the phase-space evolution of population balance equations with quadrature-based moment methods. The present version implements the quadrature method of moments (QMOM) and the conditional quadrature method of moments (CQMOM). QMOMKit is designed to be independent of any particular application and can be used either as a standalone solver for spatially homogeneous population models or as a local phase-space solver coupled to a host application. It handles the numerical aspects of the evolution of a population in phase space, including quadrature reconstruction, advection of the quadrature abscissas, and changes to the population through moment sources and phase-space boundary contributions. The number of quadrature nodes is adapted to what the supplied moments support, subject to prescribed bounds of the phase-space coordinates, and the CQMOM implementation supports an arbitrary number of phase-space dimensions as well as any conditioning order. The principal numerical algorithms of QMOMKit are verified against analytical and manufactured solutions. Standalone spatially homogeneous applications, ranging from aerosol evaporation to particle dynamics and fragmentation, are compared with Monte Carlo solutions of the same population models, while coupling examples with OpenFOAM and deal.II demonstrate how QMOMKit can be embedded in finite-volume and discontinuous-Galerkin simulation frameworks. Together, these capabilities provide a tested and reusable foundation for a wide range of population-balance problems, and a basis for extending QMOMKit to other quadrature-based moment methods.
\hfill
Repository: \url{https://github.com/polycfd/QMOMKit}
\end{abstract}

\begin{keyword}
Population balance equation \sep Quadrature-based moment methods \sep Quadrature method of moments \sep Conditional quadrature method of moments \sep Adaptive quadrature \sep Phase-space evolution
\\~\\
\textcopyright~2026. This manuscript version is made available under the CC-BY 4.0 license. \url{http://creativecommons.org/licenses/by/4.0/}
\end{keyword}

\end{frontmatter}

\begin{table}[!t]
\renewcommand{\arraystretch}{1.2}
\begin{tabular}{p{0.31\textwidth} p{0.64\textwidth}}
\hline
\textbf{Item} & \textbf{Information} \\
\hline
Program title & QMOMKit \\
Repository link & \url{https://github.com/polycfd/QMOMKit} \\
Licensing provisions & Apache 2.0 \\
Programming language & C++ \\
Nature of problem &
Population balance equations describe disperse multiphase systems through a number density function (NDF) over physical space and a phase space representing properties of the population. Directly resolving the evolution of this distribution becomes prohibitively expensive as the number of phase-space dimensions grows. Moment methods reduce this cost by evolving selected moments of the NDF, but require closure to recover information about the underlying population that is not contained in those moments.\\
Solution method &
QMOMKit provides quadrature reconstruction from raw moments using the quadrature method of moments (QMOM) and conditional quadrature method of moments (CQMOM), together with their local evolution in phase space, independently of the physical-space discretization used by a host application.
The algorithms implemented in QMOMKit reconstruct QMOM quadratures with the Wheeler algorithm and extend CQMOM to arbitrary phase-space dimensions through recursive marginal and conditional quadratures, with adaptive node counts where required. Conditional moments are obtained from weighted Vandermonde systems. Phase-space advection is advanced by moving the quadrature abscissas, while moment sources and phase-space boundary contributions advance the moments directly. Complete phase-space evolution uses Lie or Strang operator splitting, with optional realizability checks.\\
Additional comments, including restrictions and unusual features &
QMOMKit is application-independent and operates on local phase-space states. It can be used for standalone spatially homogeneous population models or embedded in host applications, while physical-space discretization, transport, parallelization, and other global solver infrastructure remain outside the library. Application-specific phase-space velocities, moment sources, and phase-space boundary contributions are supplied through callbacks. The algorithms for the reconstruction of quadratures are independent of the physical units and number density of the NDF.\\
\hline
\end{tabular}
\caption{Program summary}
\label{tab:program-summary}
\end{table}

\section{Introduction}

Many disperse multiphase systems contain large populations of particles, droplets, bubbles, aerosols, biological cells, or other discrete entities whose properties vary across the population \citep{Marchisio2013, Balachandar2024}. This polydispersity may involve size, composition, temperature, morphology, velocity, or several properties simultaneously, and can strongly influence transport, heating, phase change, reaction, aggregation, breakage, and other physical processes. A single representative value is therefore often insufficient to describe the disperse phase. Population balance equations (PBEs) provide a general framework for retaining this information \citep{Ramkrishna2000, Rigopoulos2024} by describing the population through a number density function (NDF) over physical space and a phase space characterizing the state of the population elements. The resulting equations involve mathematical operators of different nature, including hyperbolic transport and nonlocal integral operators, that together with the need to preserve properties such as moment realizability, makes their numerical solution challenging. In addition, the computational cost of directly resolving the NDF grows rapidly with increasing phase-space dimensions \citep{Shiea2020}.

Moment methods reduce the computational burden of resolving the full NDF by evolving a finite set of its moments instead \citep{Hulburt1964, Marchisio2013, Struchtrup2005}. Besides providing a compact description of the population, moments often have a direct physical interpretation and can represent quantities such as mass or momentum of the disperse phase. Their balance equations therefore provide a natural basis for conservation laws \citep{Li2022c,Passalacqua2023,Fox2024}. Moment methods have consequently found applications ranging from kinetic gas dynamics \citep{Groth2009,Torrilhon2016,Morin2026} to particle-laden flows \citep{Fan2004, Mazzei2010, Heylmun2021,Li2022c}, aerosol transport \citep{McGraw1997, Upadhyay2012, Christianson2026}, bubble columns \citep{Vikas2011a, Yuan2014, Li2019a}, cavitation \citep{Bryngelson2023}, soot formation \citep{Salenbauch2017,Xing2022, Stephens2023}, and crystallization \citep{Marchisio2026}. However, the compactness of moment-based descriptions introduces a closure problem, since the evolution of the retained moments may depend on information about the NDF that those moments do not uniquely determine. Quadrature-based moment methods (QBMM) address this problem by representing the population through a finite quadrature whose weighted sums reproduce the retained moments \citep{Marchisio2013,Li2019b}. In the foundational quadrature method of moments (QMOM) of \citet{McGraw1997}, the first $2N$ moments determine an $N$-node quadrature, originally computed using the product-difference algorithm of \citet{Gordon1968}. This quadrature approximates the NDF as a weighted sum of $N$ Dirac delta functions centered at the quadrature nodes. Consequently, integrals over the NDF can be approximated as finite weighted sums, providing estimates of the higher- or fractional-order moments needed to close moment transport equations and source terms.

A substantial body of work has extended the original QMOM formulation to address limitations that arise in different classes of population-balance problems, giving rise to a large family of QBMMs \citep{Marchisio2013, Li2019b}. For instance, the quadrature weights and abscissas are evolved directly in the direct quadrature method of moments (DQMOM) \citep{Fan2004,Marchisio2005,Attar2008,Selma2010,Mazzei2010}, avoiding repeated solution of the moment inversion problem and facilitating multivariate formulations \citep{Fox2009}, although finite-volume schemes applied to the weights and abscissas do not guarantee the conservation of the moments \citep{Li2019b}. The extended quadrature method of moments (EQMOM) replaces the Dirac delta functions of QMOM with smooth, non-negative kernel functions  \citep{Yuan2012,Madadi-Kandjani2015,Chalons2017,Pigou2018}, thereby providing an explicit continuous approximation of the NDF that is useful when the population model depends directly on the distribution itself, for example to evaluate population loss through a phase-space boundary as particles or droplets shrink to zero size.
Rather than relying on the reconstruction of the underlying distribution, the hyperbolic quadrature method of moments (HyQMOM) closes the higher-order velocity moments appearing in the free-transport flux so that the resulting moment system is globally hyperbolic and in conservative form \citep{Fox2018,Fox2022,Li2022c,Bryngelson2026,Morin2025}. 
More recently, the generalized quadrature method of moments (GQMOM) uses closures for the recurrence coefficients of the orthogonal polynomials associated with the moment set, with these closures informed by prescribed distribution families \citep{Fox2023a}. This implicitly defines the higher-order moments and permits a Gaussian quadrature with more nodes than would be available from QMOM using the same low-order moment information. 
All the QBMMs highlighted above share the broader objective of providing accurate and computationally tractable closures of finite moment systems.

Many disperse populations must be described by more than one phase-space variable because the properties of individual entities are coupled. For example, particles of different sizes may respond differently to the surrounding flow and therefore develop a size-dependent velocity distribution. In such cases, knowledge of the separate size and velocity distributions is insufficient: the population model must retain how size and velocity are correlated within the same particles. Overall, multidimensional populations are described by a joint NDF whose dependence between phase-space variables cannot, in general, be reconstructed from the corresponding marginal distributions alone.
The conditional quadrature method of moments (CQMOM), introduced by \citet{Cheng2010} and developed further by \citet{Yuan2011}, addresses this problem by expressing the multidimensional quadrature reconstruction as a hierarchy of marginal and conditional one-dimensional quadratures. The conditional distributions may vary from one branch of the hierarchy to another and can therefore represent the statistical dependence between the phase-space variables. Formulated as a recursion over the conditioning levels, this hierarchy applies to an arbitrary number of phase-space dimensions \citep{Putz2024,Bouguezzoul2026a}.
The idea of a conditional quadrature construction has subsequently been applied to hyperbolic (CHyQMOM) \citep{Fox2018,Patel2019a,Bryngelson2026} and generalized (GCQMOM) \citep{Fox2024a,Posey2026} quadrature methods of moments.
This coupling between phase-space dimensions comes, however, at a computational cost. Conditional moments must be recovered from mixed moments, successive reconstructions depend on the chosen conditioning order, and the number of conditional branches and required moments grows rapidly with the number of phase-space dimensions. Consequently, the numerical complexity of the quadrature reconstruction increases substantially as additional phase-space dimensions are introduced.

Fox and co-workers have accompanied the development of QBMMs over many years with algorithms and code snippets published alongside their research, see e.g.~\citep{Yuan2011,Marchisio2013,Nguyen2016}. These contributions have played an important role in demystifying QBMMs and enabling their independent implementation by other researchers. Despite this long-standing effort, only a relatively small number of open-source implementations are available. OpenQBMM \citep{Passalacqua2018} provides a comprehensive implementation of quadrature-based moment methods based on EQMOM, tightly integrated into OpenFOAM, making it well suited to OpenFOAM-based applications but not readily reusable as an independent numerical component in other simulation frameworks. Further examples of openly available QBMM implementations in OpenFOAM are PUFoam for the simulation of polyurethane foams \citep{Karimi2017} and an OpenQBMM-based solver for respiratory aerosols \citep{Feng2023}. Other open-source CFD codes also include different integrations of QBMMs: MFC includes a CHyQMOM implementation \citep{Wilfong2026} and Fluidity features a DQMOM implementation \citep{Davies2011}; QMOM has been integrated into MFix \citep{Fan2004, Passalacqua2010, Passalacqua2011a} and is used by TOSVA for aerosol simulations \citep{Christianson2026}. 
The open-source C++ library libMoM uses QMOM for the solution of a wide range of stochastic problems \citep{Upadhyay2012a} and
SootLib \citep{Stephens2023} is an open-source software library written in C++ that offers different methods to model soot formation, including QMOM-based algorithms. The Python-based QuadMomPy library of \citet{Putz2024} offers QBMM implementations for spatially homogeneous population balance equations and, notably, generalizes CQMOM to arbitrary dimensions.
Implemented in Mathematica and Python, QBMMLib and PyQBMMLib \citep{Bryngelson2020a} provide QMOM and HyQMOM algorithms, as well as their conditional variants, that are well suited to method development, analysis, and prototyping, but less so as performance-critical tools in large-scale compiled simulation codes. Other open-source repositories accompany specific studies, such as the EQMOM implementation in Matlab of \citet{Pigou2018}, the QBMM profiling tools in C++ of \citet{Putz2023}, or the HyQMOM implementation in Julia of \citet{Bryngelson2026}, providing excellent reference implementations while remaining specific to a particular method or purpose. Consequently, there remains a gap for an application-independent implementation of QBMMs that can be embedded in different scientific and high-performance computing frameworks, without requiring the surrounding application to adopt a particular flow solver or programming environment.

To address this gap, we present QMOMKit, a portable open-source C++ library for QBMMs, designed as an application-agnostic numerical component. QMOMKit isolates the numerical machinery required to represent and evolve a population in phase space: it reconstructs quadratures from raw moments, advances quadrature abscissas through phase-space dynamics, and accounts for changes to the population through moment sources and phase-space boundary contributions. The present version of QMOMKit implements QMOM and CQMOM, for which a reconstructed quadrature is accepted only if it satisfies explicit numerical criteria and the number of quadrature nodes is adapted to what the supplied moments support. The application supplies the physical model, while physical-space transport, spatial discretization, and global solver infrastructure remain outside the library. This separation allows the same implementation either to be used for standalone population models or to be embedded directly in existing simulation frameworks. QMOMKit is supported by a comprehensive suite of unit, regression, and verification tests, together with reproducible application examples of different dimensionality and complexity. The numerical algorithms are verified independently against analytical and manufactured solutions, and are validated in comparison to Monte Carlo reference solutions. Finally, direct couplings to OpenFOAM v14 \citep{OpenFOAMv14} and deal.II v9.8 \citep{Arndt2026} demonstrate the integration of QMOMKit into finite-volume and finite-element simulation frameworks.  Together, these verification, application, and coupling examples demonstrate that QMOMKit provides not merely a collection of quadrature algorithms, but a tested and reusable numerical foundation for applying QBMMs across different population models and computational frameworks.

The remainder of this manuscript is organized as follows. Section~\ref{sec:mathematical-foundations} introduces the population-balance and quadrature-based moment formulations underlying the present version of QMOMKit. Section~\ref{sec:architecture} describes the software architecture, followed by the principal numerical algorithms in Section~\ref{sec:numerics}, and the software interface in Section~\ref{sec:software-interface}. Important numerical components are verified in Section~\ref{sec:verification}, while Section~\ref{sec:applications} presents increasingly complex, spatially homogeneous application examples and comparisons with Monte Carlo reference solutions. Section~\ref{sec:coupling} demonstrates the coupling of QMOMKit to external finite-volume and finite-element frameworks, before the main conclusions are summarized in Section~\ref{sec:conclusions}.

\section{Mathematical foundations}
\label{sec:mathematical-foundations}

The distribution of properties across a population can be described by a number density function (NDF) $n(t,\mathbf{x},\boldsymbol{\xi})$, where $t$ denotes time, $\mathbf{x}$ is the position in physical space, and $\boldsymbol{\xi}=(\xi_1,\ldots,\xi_D)$ represents the coordinates of a $D$-dimensional phase space with support $\Omega_{\xi}\subseteq\mathbb{R}^D$. The phase-space coordinates characterize the properties of individual population elements and may represent, for example, the size and temperature of a particle, the composition of a droplet, or other properties relevant to the population under consideration. 

\subsection{Population balance and moment formulation}
\label{sec:pbe}

The evolution of an NDF $n(t,\mathbf{x},\boldsymbol{\xi})$ is generally governed by a population balance equation (PBE) \citep{Ramkrishna2000}
\begin{equation}
\frac{\partial n}{\partial t} + \nabla_{\mathbf{x}} \cdot \left(n \mathbf{u}\right) + \nabla_{\boldsymbol{\xi}} \cdot \left(n \dot{\boldsymbol{\xi}} \right) = \mathcal{S},
\label{eq:pbe}
\end{equation}
where $\mathbf{u}(t,\mathbf{x},\boldsymbol{\xi})$ is the transport velocity in physical space, $\dot{\boldsymbol{\xi}}(t,\mathbf{x},\boldsymbol{\xi})$ is the phase-space velocity, and $\mathcal{S}(t,\mathbf{x},\boldsymbol{\xi})$ is the sum of all source terms that describe processes that change the population, such as aggregation or breakage. Solving Eq.~\eqref{eq:pbe} directly requires resolving the NDF over both physical space and phase space. 

To this end, the population may be described by a finite set of the moments of the NDF \citep{McGraw1997,Marchisio2013}, which provide a compact representation of selected information about the distribution. For a multi-index $\boldsymbol{\alpha}=(\alpha_1,\ldots,\alpha_D)\in\mathbb{N}_0^D$, the corresponding phase-space monomial is defined as
\begin{equation}
\boldsymbol{\xi}^{\boldsymbol{\alpha}}
\coloneqq
\prod_{d=1}^{D}\xi_d^{\alpha_d}.
\label{eq:multi-index-monomial}
\end{equation}
Multiplying the NDF by this monomial and integrating over the phase space $\Omega_\xi$
yields the raw moments of the NDF,
\begin{equation}
M_{\boldsymbol{\alpha}}(t,\mathbf{x}) = \int_{\Omega_{\xi}} \boldsymbol{\xi}^{\boldsymbol{\alpha}} n(t,\mathbf{x},\boldsymbol{\xi}) \, \mathrm{d}\boldsymbol{\xi}.
\label{eq:raw-moment}
\end{equation}
The zeroth-order moment $M_{\boldsymbol{0}}$ represents the total number density of the population and the first-order moments are the number-density-weighted integrals of the individual phase-space coordinates. 
Higher-order moments provide the information needed to determine properties such as variance and skewness, while mixed moments contain information about the joint variation of different phase-space coordinates. 

Following the definition of the raw moments in Eq.~\eqref{eq:raw-moment}, a balance equation for $M_{\boldsymbol{\alpha}}$ is obtained by multiplying Eq.~\eqref{eq:pbe} by the monomial $\boldsymbol{\xi}^{\boldsymbol{\alpha}}$ and integrating over the phase space $\Omega_{\xi}$ \citep{Marchisio2013},
\begin{equation}
\frac{\partial M_{\boldsymbol{\alpha}}}{\partial t} + \nabla_{\mathbf{x}}\cdot \left(\int_{\Omega_{\xi}} \boldsymbol{\xi}^{\boldsymbol{\alpha}} n \mathbf{u} \, \mathrm{d}\boldsymbol{\xi}\right)
+ \int_{\Omega_{\xi}} \boldsymbol{\xi}^{\boldsymbol{\alpha}} \nabla_{\boldsymbol{\xi}}\cdot \left(n\dot{\boldsymbol{\xi}}\right) \,\mathrm{d}\boldsymbol{\xi}  = \int_{\Omega_{\xi}} \boldsymbol{\xi}^{\boldsymbol{\alpha}} \mathcal{S} \, \mathrm{d}\boldsymbol{\xi},
\label{eq:mte}
\end{equation}
where the second and third terms describe the transport of moment $M_{\boldsymbol{\alpha}}$ in physical space and phase space, respectively. Applying integration by parts to the phase-space advection term yields
\begin{equation}
\int_{\Omega_{\xi}} \boldsymbol{\xi}^{\boldsymbol{\alpha}} \nabla_{\boldsymbol{\xi}}\cdot \left(n\dot{\boldsymbol{\xi}}\right) \,\mathrm{d}\boldsymbol{\xi} = - \sum_{\substack{d=1\\\alpha_d>0}}^{D} \alpha_d \int_{\Omega_{\xi}} \boldsymbol{\xi}^{\boldsymbol{\alpha}-\boldsymbol{e}_d} \dot{\xi}_d n \,\mathrm{d}\boldsymbol{\xi} + \int_{\partial\Omega_{\xi}} \boldsymbol{\xi}^{\boldsymbol{\alpha}} n\dot{\boldsymbol{\xi}}\cdot\mathbf{n}_{\xi} \,\mathrm{d}S_{\xi} ,
\label{eq:mte-phase-space-integration-by-parts}
\end{equation}
where $\mathbf{n}_{\xi}$ is the outward unit normal on $\partial\Omega_{\xi}$ and $\boldsymbol{e}_d$ is the unit multi-index of coordinate $d$. The integral over phase space describes changes driven by the advection in phase space, whereas the boundary integral accounts for population entering or leaving the phase-space domain \citep{Fox2008,Yuan2012}. After discretization in physical space by a host numerical method and anticipating the quadrature closure introduced below, the evolution of the moment vector $\mathbf{M}_h$ associated with a host degree of freedom $h$ is governed by the semi-discrete moment equation
\begin{equation}
\frac{\mathrm{d}\mathbf{M}_h}{\mathrm{d}t} = \mathcal{A}_h(\mathbf{M}) + \mathcal{B}(\mathbf{M}_h),
\label{eq:mte-semi-discrete}
\end{equation}
where $\mathbf{M}$ denotes the collection of all local moment vectors $\mathbf{M}_h$ over the physical-space discretization. The operator $\mathcal{A}_h$ represents the discretized physical-space transport of moments and, hence, depends on the neighboring moment values contained in $\mathbf{M}$. In contrast, $\mathcal{B}$ is the local operator that describes the moment evolution in phase space.

The moment transport equation in Eq.~\eqref{eq:mte} shows that a formulation with a finite number of transported moments is generally unclosed. A physical velocity $\mathbf{u}(t,\mathbf{x},\boldsymbol{\xi})$ that depends on quantities evolving in phase space, a nonlinear phase-space velocity $\dot{\boldsymbol{\xi}}(t,\mathbf{x},\boldsymbol{\xi})$, or a moment source $\mathcal{S}(t,\mathbf{x},\boldsymbol{\xi})$ may introduce integrals that cannot be determined uniquely from the transported moments, because moments of fractional or higher order are required. Consequently, additional information about the underlying distribution is required to close Eq.~\eqref{eq:mte}. QBMMs, such as QMOM and CQMOM, address this closure problem by constructing a finite set of weighted quadrature nodes that reproduces the transported low-order moments and can approximate high- and fractional-order moments \citep{Marchisio2013}.

\subsection{Quadrature method of moments}

Let us first consider a population described by a single phase-space coordinate $\xi$. The quadrature method of moments (QMOM) of \citet{McGraw1997} represents the moment information by a set of $N$ quadrature nodes,
\begin{equation}
n(\xi) \approx \sum_{i=1}^{N} w_i\,\delta(\xi-\xi_i),
\label{eq:qmom-discrete}
\end{equation}
where each quadrature node consists of an abscissa $\xi_i\in\Omega_{\xi}$ and an associated weight $w_i > 0$, and $\delta$ denotes the Dirac delta function. The weights retain the normalization of the NDF, such that
\begin{equation}
M_0 = \sum_{i=1}^{N}w_i \label{eq:qmom-M0}
\end{equation}
is the total number density represented by the quadrature.  An $N$-node quadrature contains $2N$ unknowns, which QMOM determines by requiring the quadrature to reproduce the first $2N$ raw moments \citep{McGraw1997,Marchisio2003},
\begin{equation}
M_\ell = \sum_{i=1}^{N} w_i\xi_i^\ell, \qquad \ell=0,\ldots,2N-1.
\label{eq:qmom-moment-matching}
\end{equation}
The most commonly used algorithms to solve this moment-inversion problem are based on either the product-difference algorithm of \citet{Gordon1968} or the algorithm of \citet{Wheeler1974}.

Once the abscissas and weights have been determined, the quadrature defines reconstructed moments of arbitrary (integer or fractional) order $\ell$,
\begin{equation}
\widehat{M}_\ell = \sum_{i=1}^{N} w_i\xi_i^\ell.
\label{eq:qmom-moment-approximation}
\end{equation}
For integer orders $0\leq \ell < 2N$, the reconstructed moments coincide with the moments imposed in Eq.~\eqref{eq:qmom-moment-matching}. Moments of higher or fractional order are not prescribed independently and, instead, their values are approximated by reconstructing them using the quadrature. More generally, a phase-space integral with a sufficiently regular integrand $g$ is approximated as
\begin{equation}
\int_{\Omega_{\xi}}
g(\xi)n(\xi)\,\mathrm{d}\xi
\approx
\sum_{i=1}^{N}
w_i g(\xi_i),
\label{eq:qmom-integral-closure}
\end{equation}
which provides the basic closure relation of QMOM. The accuracy of this approximation depends on the number of quadrature nodes, the retained moments, and the variation of $g$ over the support of the distribution \citep{Gautschi2004, Grosch2007}. Once the quadrature has been reconstructed from the transported low-order moments, integrals over the NDF are replaced by finite sums over the quadrature nodes, with the integrand evaluated at each node's abscissa and multiplied by the corresponding weight. This makes it possible to evaluate terms that depend on the NDF without reconstructing it. 

\subsection{Conditional quadrature method of moments}
\label{sec:cqmom}

The conditional quadrature method of moments (CQMOM) \citep{Cheng2010,Yuan2011} extends QMOM to populations described by several, potentially correlated, phase-space dimensions. Its construction is most easily introduced for a two-dimensional phase space $\boldsymbol{\xi}=(\xi_1,\xi_2)$. The joint NDF may be written as
\begin{equation}
n(\xi_1,\xi_2) = n_1(\xi_1)\, p_{2|1}(\xi_2\mid\xi_1),
\label{eq:conditional-factorization}
\end{equation}
where
\begin{equation}
n_1(\xi_1) = \int_{\Omega_{\xi_2}} n(\xi_1,\xi_2)\,\mathrm{d}\xi_2
\end{equation}
is the marginal NDF of $\xi_1$.
For $n_1(\xi_1)>0$, the conditional density is
\begin{equation}
p_{2|1}(\xi_2\mid\xi_1) = \frac{n(\xi_1,\xi_2)}{n_1(\xi_1)},
\end{equation}
and therefore satisfies
\begin{equation}
\int_{\Omega_{\xi_2}} p_{2|1}(\xi_2\mid\xi_1)\,\mathrm{d}\xi_2 = 1.
\end{equation}

The first step in reconstructing a multidimensional quadrature with CQMOM is a QMOM reconstruction of the quadrature for the marginal distribution from the pure moments $M_{p,0}$, where the subscripts refer to the moment order in the corresponding phase-space dimension, such that
\begin{equation}
n_1(\xi_1) = \sum_{i_1=1}^{N_1} w^{(1)}_{i_1} \delta\!\left(\xi_1-\xi^{(1)}_{i_1}\right),
\label{eq:cqmom-marginal-quadrature}
\end{equation}
where
\begin{equation}
M_{0,0} = \sum_{i_1=1}^{N_1}w^{(1)}_{i_1}.
\end{equation}
CQMOM next constructs a separate quadrature for the distribution of $\xi_2$ conditioned on each marginal abscissa $\xi^{(1)}_{i_1}$. The conditional moments required for these reconstructions are not available directly. 
For the continuous distribution $p_{2|1}$, the $q$th conditional moment of $\xi_2$ at a given value of $\xi_1$ is defined as
\begin{equation}
\mu_q^{(2|1)}(\xi_1) = \int_{\Omega_{\xi_2}} \xi_2^q \, p_{2|1}(\xi_2\mid\xi_1)\,\mathrm{d}\xi_2.
\label{eq:cqmom-conditional-moment}
\end{equation}
The mixed raw moments then satisfy
\begin{equation}
M_{p,q} = \int_{\Omega_{\xi_1}} \xi_1^p \ n_1(\xi_1) \ \mu_q^{(2|1)}(\xi_1) \ \mathrm{d}\xi_1.
\end{equation}
CQMOM represents this integral using the marginal quadrature and determines branch-specific conditional moments $\mu^{(2|1)}_{q|i_1}$ such that
\begin{equation}
M_{p,q} = \sum_{i_1=1}^{N_1} w^{(1)}_{i_1} \left(\xi^{(1)}_{i_1}\right)^p \mu^{(2|1)}_{q|i_1}.
\label{eq:cqmom-conditional-moment-relation}
\end{equation}
For each fixed value of $q$, the $N_1$ conditional moments $\mu^{(2|1)}_{q|i_1}$ are obtained by taking $p=0,\ldots,N_1-1$, defining the linear system
\begin{equation}
\mathbf{V}^{(1)} \mathbf{W}^{(1)} \boldsymbol{\mu}^{(q)} = \mathbf{m}^{(q)},
\label{eq:cqmom-conditional-moment-system}
\end{equation}
with the Vandermonde matrix
\begin{equation}
\mathbf{V}^{(1)} =
\begin{pmatrix}
1 & 1 & \cdots & 1 \\
\xi^{(1)}_1 & \xi^{(1)}_2 & \cdots & \xi^{(1)}_{N_1} \\
\vdots & \vdots & \ddots & \vdots \\
\left(\xi^{(1)}_1\right)^{N_1-1}
&
\left(\xi^{(1)}_2\right)^{N_1-1}
&
\cdots
&
\left(\xi^{(1)}_{N_1}\right)^{N_1-1}
\end{pmatrix},
\end{equation}
the diagonal weight matrix
\begin{equation}
\mathbf{W}^{(1)}
=
\operatorname{diag}
\left(
w^{(1)}_1,\ldots,w^{(1)}_{N_1}
\right),
\end{equation}
the conditional moment vector
\begin{equation}
\boldsymbol{\mu}^{(q)}
=
\begin{pmatrix}
\mu^{(2|1)}_{q|1} \\
\vdots \\
\mu^{(2|1)}_{q|N_1}
\end{pmatrix},
\end{equation}
and the mixed raw moment vector
\begin{equation}
\mathbf{m}^{(q)}
=
\begin{pmatrix}
M_{0,q} \\
\vdots \\
M_{N_1-1,q}
\end{pmatrix}.
\end{equation}

Solving Eq.~\eqref{eq:cqmom-conditional-moment-system} yields the discrete conditional moment $\mu^{(2|1)}_{q|i_1}$ associated with each marginal node $i_1$. The coefficient matrix $\mathbf{V}^{(1)}\mathbf{W}^{(1)}$ depends only on the marginal quadrature: $\mathbf{V}^{(1)}$ is formed from the marginal abscissas and multiplication by $\mathbf{W}^{(1)}$ scales its columns by the corresponding marginal weights. The same matrix is used for every required power $q$ of $\xi_2$ and only the mixed-moment vector $\mathbf{m}^{(q)}$ on the right-hand side changes. The factorization of the weighted Vandermonde matrix can therefore be reused for all values of $q$. If the marginal abscissas are distinct and the marginal weights are nonzero, the coefficient matrix is non-singular and these conditional moments are uniquely determined.

Based on the recovered conditional moments, each marginal node $i_1$ is associated with a one-dimensional conditional moment sequence. A separate QMOM reconstruction is applied to each sequence,
\begin{equation}
p_{2|1}\!\left(\xi_2\mid\xi^{(1)}_{i_1}\right) = \sum_{i_2=1}^{N_{2|i_1}} w^{(2|1)}_{i_2|i_1} \delta\!\left(\xi_2-\xi^{(2|1)}_{i_2|i_1}\right),
\label{eq:cqmom-conditional-quadrature}
\end{equation}
where
\begin{equation}
\sum_{i_2=1}^{N_{2|i_1}} w^{(2|1)}_{i_2|i_1} = 1.
\end{equation}
Because the conditional abscissas and weights may vary between marginal nodes, CQMOM can represent a dependence between $\xi_1$ and $\xi_2$, and, thus, does not assume that the joint distribution is the product of two independent marginal distributions. Combining the marginal and conditional quadratures gives
\begin{equation}
n(\xi_1,\xi_2) = \sum_{i_1=1}^{N_1} \sum_{i_2=1}^{N_{2|i_1}} W_{i_1,i_2} \delta\!\left(\xi_1-\xi^{(1)}_{i_1}\right) \delta\!\left(\xi_2-\xi^{(2|1)}_{i_2|i_1} \right),
\label{eq:cqmom-joint-quadrature}
\end{equation}
where the joint weight is $W_{i_1,i_2} = w^{(1)}_{i_1} w^{(2|1)}_{i_2|i_1}$. The corresponding mixed moments are approximated by
\begin{equation}
M_{p,q} \approx \sum_{i_1=1}^{N_1} \sum_{i_2=1}^{N_{2|i_1}} W_{i_1,i_2} \left(\xi^{(1)}_{i_1}\right)^{p} \left(\xi^{(2|1)}_{i_2|i_1}\right)^{q}.
\label{eq:cqmom-mixed-moment}
\end{equation}

In summary, CQMOM replaces a multidimensional moment-inversion problem by a hierarchy of one-dimensional inversions based on conditional moment sets. Each quadrature node at one level may carry its own conditional quadrature at the next level, producing a tree-like representation of the joint distribution. In this manner, the CQMOM methodology for two dimensions shown above can be readily extended to a phase space of arbitrary dimensions, as discussed in Section \ref{sec:numerics-cqmom}.

\section{QMOMKit architecture}
\label{sec:architecture}

QMOMKit is designed as an application-independent numerical library for quadrature-based moment methods. Its design follows from the semi-discrete moment transport equation shown in Eq.~\eqref{eq:mte-semi-discrete}, in which physical-space transport and phase-space evolution appear as distinct operators, $\mathcal{A}_h(\mathbf{M})$ and $\mathcal{B}(\mathbf{M}_h)$, that can be advanced separately. QMOMKit provides the phase-space operator $\mathcal{B}$. The central design principle of the library is therefore to treat phase-space evolution as a local operation on moment states, independent of how those moments are discretized and transported in physical space. This allows QMOMKit to be used either as a standalone solver for the spatially uniform evolution of a population, as shown in Section~\ref{sec:applications}, or coupled to a host application that transports the moments in physical space, as demonstrated in Section~\ref{sec:coupling}. In the latter case, the host advances the moments through $\mathcal{A}_h$ and remains responsible for mesh handling, the physical-space discretization, the boundary conditions of the physical domain, parallel communication, and global time integration, while QMOMKit advances the local moment state $\mathbf{M}_h$ of the host degree of freedom $h$ through $\mathcal{B}$. Within this local operation, the raw moments are the primary variables, since they are the quantities that the host transports and that the moment transport equations govern. Whenever required, QMOMKit reconstructs a quadrature locally from $\mathbf{M}_h$ using the prescribed numerical options, such as the maximum number of nodes. Once a quadrature is accepted, the moments are recomputed from it, so that the moments and their quadrature remain a coherent numerical unit rather than two independently evolving representations of the population. 

The phase-space processes themselves are advanced in the representation that suits them, as discussed in Section \ref{sec:moment-evolution}. Phase-space advection moves existing members of the population and is advanced by moving the quadrature abscissas along characteristics in phase space, after which the moments are evaluated from the updated quadrature. Source terms that change the population, including moment sources and exchange at the boundary of phase space, cannot be represented by moving abscissas. Instead, the moments are advanced directly, after which the quadrature is reconstructed. The two treatments are combined by operator splitting.
What drives these processes is defined by the application rather than by QMOMKit: The standalone or host application defines the physical meaning and ordering of the phase-space coordinates and supplies the phase-space velocity, the moment sources, and the boundary contributions, which QMOMKit evaluates during the phase-space evolution. Fixed numerical choices, such as the phase-space dimension and the reconstruction policies are defined by the configuration, while the runtime state contains the evolving moments and quadrature. This separation keeps the physical model independent of the numerical algorithms used to represent and evolve the population, so that the same algorithms can serve different applications.

Because QMOMKit is intended to be embedded in other codes, its users need to be able to rely on its numerical behavior without inspecting the implementation. Following the principle that confidence in numerical software should be established through reproducible evidence \citep{Barba2026}, QMOMKit includes a test suite that, at the time of writing this manuscript, comprises 986 unique tests: 710 unit tests isolate individual operations and software contracts, 229 verification tests compare numerical results with analytical, manufactured, or independently established reference solutions, 46 regression tests preserve previously established behavior, and a public-header test verifies that the library's public header files can be included and compiled independently by external users.

The following sections build on these design principles. Section~\ref{sec:numerics} describes the numerical algorithms behind the phase-space operator, Section~\ref{sec:software-interface} the software interface through which applications supply their models and advance the moment states, and Section~\ref{sec:verification} a selection of verification studies.

\section{Numerical algorithms}
\label{sec:numerics}

The numerical core of QMOMKit comprises the reconstruction and evolution of quadrature-based moment representations. This section first describes the reconstruction of a quadrature from a moment set in Section~\ref{sec:numerics-quadrature}, followed by the reconstruction of a multidimensional quadrature in Section~\ref{sec:numerics-cqmom}. Section~\ref{sec:moment-evolution} then describes how the moments evolve in phase space, with continuous motion advanced by advecting the quadrature and changes of the population advanced directly through the moments. Section~\ref{sec:node-count-selection} explains how the active node count is selected as the moments evolve and Section~\ref{sec:moment-realizability} how the realizability of the evolving moments can be checked.

\subsection{Quadrature reconstruction}
\label{sec:numerics-quadrature}

The basic operation of QMOMKit is the reconstruction of a one-dimensional quadrature from a moment set, which is used directly in QMOM as well as for the reconstruction of every marginal and conditional quadrature in CQMOM. Given the moments $M_0,\ldots,M_{2N-1}$ of a distribution $n(\xi)$, the $N$ abscissas of the quadrature in Eq.~\eqref{eq:qmom-discrete} are the roots of the monic polynomial $P_N$ of degree $N$ that is orthogonal with respect to $n$ \citep{Gautschi1970,Press1990,Gautschi2004}. These orthogonal polynomials satisfy the three-term recurrence
\begin{equation}
P_{k+1}(\xi) = (\xi-a_k)\,P_k(\xi) - b_k\,P_{k-1}(\xi), \qquad P_{-1}=0, \quad P_0=1.
\label{eq:orthogonal-polynomial-recurrence}
\end{equation}
The recurrence coefficients $a$ and $b$ define the symmetric tridiagonal Jacobi matrix
\begin{equation}
\mathbf{J}_N =
\begin{pmatrix}
a_0 & \sqrt{b_1} & & \\
\sqrt{b_1} & a_1 & \ddots & \\
 & \ddots & \ddots & \sqrt{b_{N-1}} \\
 & & \sqrt{b_{N-1}} & a_{N-1}
\end{pmatrix},
\label{eq:jacobi-matrix}
\end{equation}
the eigenvalues of which are the abscissas $\xi_i$, and each weight $w_i$ is $M_0$ times the squared first component of the corresponding normalized eigenvector \citep{Golub1969}. QMOMKit solves this eigenvalue problem with the Jacobi rotation method \citep{Golub2013}. Before the rotations, the matrix is multiplied by the power of two that brings its largest entry into the interval $[1,2)$, and the computed eigenvalues are divided by the same factor afterwards. Since a multiplication by a power of two is exact in floating-point arithmetic \citep{Higham2002}, this scaling introduces no rounding error, but it makes the convergence criterion, which compares the off-diagonal entries with the machine precision $\epsilon_{\mathrm{mach}}$, relative to the magnitude of the matrix. The resulting quadrature $\mathcal{Q}_N=\{(\xi_i,w_i)\}_{i=1}^{N}$ reproduces the moments $M_0,\ldots,M_{2N-1}$ exactly.

QMOMKit computes the recurrence coefficients with the algorithm of \citet{Wheeler1974}, a variant of the long quotient-modified difference algorithm of \citet{Sack1971}. The algorithm is based on the quantities \citep{Wheeler1974,Marchisio2013}
\begin{equation}
\sigma_{k,\ell} = \int_{\Omega_{\xi}} P_k(\xi)\,\xi^{\ell}\,n(\xi)\,\mathrm{d}\xi,
\label{eq:wheeler-sigma}
\end{equation}
which start from $\sigma_{-1,\ell}=0$ and $\sigma_{0,\ell}=M_{\ell}$ and are obtained for increasing $k$ from
\begin{equation}
\sigma_{k,\ell} = \sigma_{k-1,\ell+1} - a_{k-1}\,\sigma_{k-1,\ell} - b_{k-1}\,\sigma_{k-2,\ell}
\label{eq:wheeler-sigma-recurrence}
\end{equation}
for $\ell=k,\ldots,2N-k-1$, together with $a_0=M_1/M_0$, $b_0=0$, and
\begin{equation}
a_k = \frac{\sigma_{k,k+1}}{\sigma_{k,k}} - \frac{\sigma_{k-1,k}}{\sigma_{k-1,k-1}},
\qquad
b_k = \frac{\sigma_{k,k}}{\sigma_{k-1,k-1}},
\qquad k=1,\ldots,N-1.
\label{eq:wheeler-coefficients}
\end{equation}
Comparative studies \citep{Upadhyay2012, John2012} found this type of algorithm to be more robust and more efficient than the product-difference algorithm \citep{Gordon1968} used in the original QMOM \citep{McGraw1997}.

The construction above assumes that the moments determine a quadrature with $N$ distinct nodes, which is not always the case. Since $P_k$ is orthogonal to all polynomials of lower degree, $\sigma_{k,k}=\int_{\Omega_{\xi}}P_k^2\,n\,\mathrm{d}\xi$. This quantity is positive if the support of $n$ contains more than $k$ points. If $n$ is instead a $k$-point distribution, that is, a distribution of $k$ weighted points at distinct abscissas,
\begin{equation}
n(\xi) = \sum_{i=1}^{k} w_i\,\delta(\xi-\xi_i), \qquad w_i>0,
\label{eq:k-point-distribution}
\end{equation}
then $P_k$ vanishes at all $k$ abscissas and $\sigma_{k,k}=0$ \citep{Gautschi2004}. The moments of such a distribution lie on the boundary of the realizable set and determine the distribution uniquely \citep{Schmudgen2017,Laurent2017}. The recurrence then breaks down, since $b_k=0$ and $a_k$ is undefined. Distributions of this kind occur, for example, when particles are formed by nucleation at a single size \citep{Nguyen2016}, when all elements of a population relax towards the same state, such that its moments are numerically indistinguishable from those of a $k$-point distribution, or when a quadrature weight approaches zero \citep{Grosch2007,Putz2023}. For this reason, \citet{Yuan2011} proposed an adaptive form of the Wheeler algorithm. Given a configured maximum number of nodes $N^{\max}$, it limits the number of nodes to the first $k < N^{\max}$, counting upwards from $k=1$, for which $\sigma_{k,k}\leq 0$, or to $N^{\max}$ if no such $k$ exists. In addition, it rejects a quadrature whose smallest weight is too small compared with its largest weight, or whose abscissas lie too close to each other, and reduces the number of nodes until a quadrature passes both checks. Similarly, \citet{Nguyen2016} detected moment sets on the boundary of the realizable set from the recurrence coefficients rather than from Hankel determinants.

QMOMKit follows the same adaptive procedure but decides differently when $\sigma_{k,k}$ vanishes. In floating-point arithmetic, $\sigma_{k,k}$ of a $k$-point distribution is not exactly zero. It results from the cancellation of much larger terms in Eq.~\eqref{eq:wheeler-sigma-recurrence} and retains a rounding error of either sign. If this rounding error is negative, the condition $\sigma_{k,k}\leq 0$ is met and the node count is correctly limited to $k$. If it is positive, however, the condition is not met and the algorithm constructs a quadrature with more nodes than the distribution supports. \citet{Laurent2017} addressed this with a fixed threshold, which depends on the unit of $\xi$. QMOMKit instead estimates the size of these rounding errors directly, in the manner of a running error analysis \citep{Higham2002}. Evaluating Eq.~\eqref{eq:wheeler-sigma-recurrence} with absolute values,
\begin{equation}
\bar{\sigma}_{k,\ell} = \bar{\sigma}_{k-1,\ell+1} + |a_{k-1}|\,\bar{\sigma}_{k-1,\ell} + |b_{k-1}|\,\bar{\sigma}_{k-2,\ell},
\qquad
\bar{\sigma}_{-1,\ell}=0,
\quad
\bar{\sigma}_{0,\ell}=|M_{\ell}|,
\label{eq:wheeler-sigma-bound}
\end{equation}
gives the magnitude of the terms whose cancellation produces $\sigma_{k,\ell}$, so that a value of $\sigma_{k,k}$ below a small multiple of $\epsilon_{\mathrm{mach}}\,\bar{\sigma}_{k,k}$ cannot be distinguished from zero. QMOMKit therefore determines the number of nodes $N$ supported by the moments with the same rule, but replaces the condition $\sigma_{k,k}\leq 0$ by
\begin{equation}
\sigma_{k,k} \leq C_{\sigma}\,\epsilon_{\mathrm{mach}}\,\bar{\sigma}_{k,k}.
\label{eq:wheeler-degeneracy-test}
\end{equation}
The safety factor is $C_{\sigma}=100$.

For this $N$, QMOMKit constructs the corresponding candidate quadrature $\mathcal{Q}_N$, which is accepted if
\begin{align}
\rho_w &= \frac{\min_i w_i}{\max_i w_i} > \varepsilon_w,
\label{eq:quadrature-acceptance-criteria-w}
\\
\rho_{\xi} &= \frac{\min_{i\neq j}|\xi_i-\xi_j|}{s_{\xi}} > \varepsilon_{\xi},
\label{eq:quadrature-acceptance-criteria-xi}
\end{align}
and
\begin{align}
\xi_i &\in \Omega_{\xi}\ \ \text{for } i=1,\ldots,N,
\label{eq:quadrature-acceptance-criteria-Omega}
\end{align}
where $\Omega_{\xi}$ is a prescribed bounded support of the coordinate, or $\Omega_{\xi}=\mathbb{R}$ if none is prescribed. The weight ratio $\rho_w$ and the separation ratio $\rho_{\xi}$ apply for $N>1$, with the separation measured relative to
\begin{equation}
s_{\xi} =
\begin{cases}
\max_i |\xi_i|, & N=2,\\[1mm]
\max_{i\neq j}|\xi_i-\xi_j|, & N>2.
\end{cases}
\label{eq:quadrature-separation-scale}
\end{equation}
If the candidate quadrature is rejected, $N$ is reduced by one and a new candidate is constructed, if necessary down to a single node. Because the degeneracy test already removes the nodes that the moments do not support, these criteria only need to exclude nearly singular candidates, and QMOMKit uses the small default thresholds $\varepsilon_w=\varepsilon_{\xi}=10^{-10}$. The support condition rejects candidates that leave $\Omega_{\xi}$ because of rounding errors or spurious moments, since the quadrature nodes of moments that are realizable on a bounded interval lie within that interval \citep{Gautschi2004}. 

An important consequence of these choices is that the quadrature reconstruction does not depend on the unit of the phase-space coordinate or on the number density. Under a change of unit $\xi\to L\xi$ and a scaling $n\to\Lambda n$, with constants $L>0$ and $\Lambda>0$, the moments transform as $M_{\ell}\to\Lambda L^{\ell}M_{\ell}$, the recurrence coefficients as $a_k\to L\,a_k$ and $b_k\to L^2 b_k$, and both $\sigma_{k,\ell}$ and $\bar{\sigma}_{k,\ell}$ as $\Lambda L^{k+\ell}$, so that the degeneracy test in Eq.~\eqref{eq:wheeler-degeneracy-test} gives the same node count. Since $a_k\to L\,a_k$ and $b_k\to L^2 b_k$, every entry of the Jacobi matrix in Eq.~\eqref{eq:jacobi-matrix} is multiplied by $L$, while $\Lambda$ cancels in the ratios that define $a_k$ and $b_k$. The eigenvalues of $L\,\mathbf{J}_N$ are those of $\mathbf{J}_N$ multiplied by $L$ and its eigenvectors are unchanged, so that the abscissas scale with $L$ and the weights, through $M_0$, with $\Lambda$. The power-of-two scaling of the Jacobi iteration preserves this property in floating-point arithmetic. Without this scaling, the convergence criterion would be absolute: for coordinates with small values, the iteration could terminate before the matrix is sufficiently diagonalized, while for large values it might fail to satisfy the convergence criterion. A verification test in the QMOMKit test suite confirms that quadratures are recovered with the same relative accuracy for coordinate values scaled by factors between $10^{-20}$ and $10^{20}$. The criteria in Eqs.~\eqref{eq:quadrature-acceptance-criteria-w}--\eqref{eq:quadrature-acceptance-criteria-Omega} are dimensionless and the support check uses a tolerance relative to the magnitude of the bounds. Apart from rounding, the reconstructed quadrature therefore simply follows a change of unit and number density. The only exception is a lower limit on $M_0$, by default $10^{-30}$, below which no quadrature is reconstructed and the moment set is treated as \textit{empty}. Apart from this limit, the moments do not have to be rescaled or non-dimensionalized to improve the conditioning and stability of the quadrature reconstruction \citep{Grosch2007,Plath2024} or expressed in terms of the volume fraction instead of the number density \citep{Heylmun2021}.

Adaptive quadrature reconstruction is the default in QMOMKit. If the user chooses the non-adaptive mode, the configured node count is used as given and the weight and separation criteria are not applied. The reconstruction then fails if the degeneracy test in Eq.~\eqref{eq:wheeler-degeneracy-test} shows that the moments support fewer nodes than configured or if the quadrature violates a prescribed support.

\subsection{Conditional quadrature}
\label{sec:numerics-cqmom}

For a multivariate NDF, QMOMKit reconstructs the quadrature with CQMOM, which replaces the multidimensional moment inversion by the hierarchy of one-dimensional inversions introduced in Section~\ref{sec:cqmom} \citep{Yuan2011}. Following \citet{Putz2024} and \citet{Bouguezzoul2026a}, QMOMKit applies this construction recursively, so that the same implementation handles any number of phase-space dimensions and any conditioning order without dimension-specific code.

Two orderings of the phase-space coordinates have to be distinguished. The \textit{phase-space order} $\boldsymbol{\xi}=(\xi_1,\ldots,\xi_D)$ is the order in which the application defines the coordinates. Moments, quadrature nodes, and configuration data always refer to this order, and a raw moment $M_{\boldsymbol{\alpha}}$ is indexed by the multi-index $\boldsymbol{\alpha}\in\mathbb{N}_0^D$ in this order. The \textit{conditioning order} is the order in which CQMOM reconstructs a multidimensional quadrature across the phase-space coordinates. It is given by the permutation
\begin{equation}
\boldsymbol{\pi} = (\pi_1,\ldots,\pi_D) \in S_D,
\label{eq:qmomkit-conditioning-order}
\end{equation}
where $S_D$ denotes the set of all permutations of $(1,\ldots,D)$. CQMOM first reconstructs the marginal quadrature of $\xi_{\pi_1}$, then the quadratures of $\xi_{\pi_2}$ conditioned on the nodes of $\xi_{\pi_1}$, and so on. Changing the conditioning order changes the conditional systems that are solved and the resulting quadrature \citep{Yuan2011,Li2022c}, and can decide whether the recovered conditional moments admit the configured number of nodes \citep{Favero2014}. It does not, however, change the meaning of the coordinates or of the moments.

At the first conditioning level, the marginal quadrature of $\xi_{\pi_1}$ is reconstructed from the pure moments of $\xi_{\pi_1}$ with the algorithm of Section~\ref{sec:numerics-quadrature}. At each subsequent level $m$, every node of the quadrature constructed so far carries its own conditional quadrature for $\xi_{\pi_m}$. The conditional moments of this quadrature are obtained from the mixed moments by solving weighted Vandermonde systems of the form of Eq.~\eqref{eq:cqmom-conditional-moment-system}, whose coefficient matrices are built from the abscissas and weights of the preceding levels, and whose right-hand sides contain the mixed moments of increasing order in $\xi_{\pi_m}$. Each set of conditional moments is then inverted with the algorithm of Section~\ref{sec:numerics-quadrature}. The result is a tree of quadratures. A node at level $m$ is identified by the index vector
\begin{equation}
\mathbf{i}_{1:m} = (i_1,\ldots,i_m), \qquad 1 \leq i_m \leq N_{m,\mathbf{i}_{1:m-1}},
\label{eq:qmomkit-branch-index}
\end{equation}
where $i_j$ numbers the nodes at level $j$ along the branch and $N_{m,\mathbf{i}_{1:m-1}}$ is the number of nodes that the branch $\mathbf{i}_{1:m-1}$ carries at level $m$. A complete branch $\mathbf{i}=(i_1,\ldots,i_D)$ defines a terminal node of the multidimensional quadrature. Its abscissa collects the abscissas $\xi_{\pi_m}^{(\mathbf{i}_{1:m})}$ encountered along the branch, reordered into the phase-space order, and its weight is the product of the marginal weight and the conditional weights along the branch,
\begin{equation}
W_{\mathbf{i}} = w^{(1)}_{i_1} \prod_{m=2}^{D} w^{(m)}_{i_m|\mathbf{i}_{1:m-1}}.
\label{eq:qmomkit-terminal-weight}
\end{equation}
The marginal weights carry the number density, while the conditional weights distribute it among the branches. With $\mathcal{T}_D$ denoting the set of terminal quadrature nodes, a raw moment is approximated by
\begin{equation}
M_{\boldsymbol{\alpha}} \approx \sum_{\mathbf{i}\in\mathcal{T}_D} W_{\mathbf{i}} \prod_{m=1}^{D}
\left(\xi_{\pi_m}^{(\mathbf{i}_{1:m})}\right)^{\alpha_{\pi_m}}.
\label{eq:qmomkit-multidimensional-moment}
\end{equation}
For $D=2$ and $\boldsymbol{\pi}=(1,2)$, this expression reduces to \eqref{eq:cqmom-mixed-moment}.

CQMOM does not use all raw moments up to the highest order in every coordinate. The marginal quadrature of $\xi_{\pi_1}$ with $N^{\max}$ nodes requires the $2N^{\max}$ pure moments of this coordinate. Each subsequent level requires mixed moments up to order $2N^{\max}-1$ in the coordinate reconstructed at that level, but only up to order $N^{\max}-1$ in each coordinate reconstructed before, because these orders suffice to determine the conditional moments on the $N^{\max}$ nodes of that coordinate. The moments used in this way form the \textit{required moment set}. Their number grows much more slowly with the phase-space dimension than the number of all raw moments up to order $2N^{\max}-1$ in each coordinate: for five coordinates with three nodes each, CQMOM requires 606 moments instead of 7776 (see Table~\ref{tab:verification-cqmom-affine}). QMOMKit stores and evolves only the required moments, which are therefore the moments that a host application transports in physical space. Which mixed moments are required depends on the conditioning order.

By default, the weighted Vandermonde systems that define the conditional moments, as defined in Eq.~\eqref{eq:cqmom-conditional-moment-system}, are solved in QMOMKit with the Björck-Pereyra algorithm \citep{Bjorck1970,Higham1987,Press1992}, which exploits the structure of the Vandermonde matrix and requires $\mathcal{O}(N^2)$ operations for each right-hand side. Alternatively, QMOMKit also offers general solvers based on Gaussian elimination, Householder QR factorization with column pivoting, and singular-value decomposition, each with a cost of $\mathcal{O}(N^3)$ \citep{Press1992,Golub2013}. All of these solvers reject numerically singular systems, the Björck-Pereyra solver on the basis of a condition-number estimate obtained from the bound of \citet{Gautschi1962} on the norm of the inverse Vandermonde matrix, so that the solver choice affects only the rounding errors of non-singular systems. Also available in QMOMKit is a solver using the Moore-Penrose pseudoinverse from a truncated singular-value decomposition, in which singular values below a relative cutoff are discarded \citep{Golub2013, Marco2019}. Unlike the other solvers, it returns the minimum-norm least-squares solution for rank-deficient systems and can therefore provide conditional moments where the other solvers report a singular system.

The conditional moments are themselves results of these solves and therefore contain rounding errors. The size of these errors is set by the typical magnitude of the values of $\xi_{\pi_m}$ in the population, measured by the root-mean-square value 
\begin{equation}
    \xi_{\pi_m}^{\mathrm{rms}} = \left(\frac{M_{2\boldsymbol{e}_{\pi_m}}}{M_{\boldsymbol{0}}}\right)^{1/2},
\end{equation}
rather than by the conditional moments of the branch. On a branch where the conditional distribution is concentrated at a single point, these rounding errors appear as a small spurious spread. The bound $\bar{\sigma}_{k,k}$ in Eq.~\eqref{eq:wheeler-degeneracy-test}, which is computed from the conditional moments alone, does not account for these rounding errors, so the degeneracy test may retain additional nodes. Instead of iteratively refining the solution of the Vandermonde systems to reduce these rounding errors \citep{Yuan2011}, QMOMKit accounts for them in the adaptive mode by replacing $\bar{\sigma}_{k,k}$ for conditional moment sets by $\max\{\bar{\sigma}_{k,k},\,\mu_0\,(\xi_{\pi_m}^{\mathrm{rms}})^{2k}\}$, such that
\begin{equation}
\sigma_{k,k} \leq C_{\sigma}\,\epsilon_{\mathrm{mach}}\,\max\left\{\bar{\sigma}_{k,k},\,\mu_0\left(\frac{M_{2\boldsymbol{e}_{\pi_m}}}{M_{\boldsymbol{0}}}\right)^{k}\right\},
\label{eq:cqmom-degeneracy-test}
\end{equation}
where $\mu_0$ is the zeroth conditional moment, which equals one up to rounding errors. This lower limit scales in the same way as $\sigma_{k,k}$ when the unit of $\xi_{\pi_m}$ or the number density changes, so the reconstruction remains independent of both.

Because each set of conditional moments is inverted separately, the number of nodes may differ from branch to branch. In practice, the conditional moments do not always admit the configured number of nodes \citep{Favero2014,Heylmun2021}. Different branches may represent conditional distributions of very different shape and a branch whose conditional distribution is concentrated at a few points then carries fewer nodes than its neighbors. For each coordinate $\xi_d$, the configured maximum $N_d^{\max}$ is an upper limit for the node counts $N_{m,\mathbf{i}_{1:m-1}}$ of all branches at the conditioning level $m$ with $\pi_m=d$. The branches present after a reconstruction, together with their node counts, form the \textit{active quadrature topology}, on which all subsequent branch-wise operations are carried out. In the non-adaptive mode, all branches at the same level carry the configured number of nodes. How the active node counts are selected as the moments evolve in time is described in Section~\ref{sec:node-count-selection}.

\subsection{Moment evolution in phase space}
\label{sec:moment-evolution}

Following Eqs.~\eqref{eq:mte}--\eqref{eq:mte-semi-discrete}, the local evolution of a moment set in phase space, described by the operator $\mathcal{B}$, is governed by
\begin{equation}
    \frac{\mathrm{d}{M}_{\boldsymbol{\alpha}}}{\mathrm{d}t} =  \sum_{\substack{d=1\\\alpha_d>0}}^{D} \alpha_d \int_{\Omega_{\xi}} \boldsymbol{\xi}^{\boldsymbol{\alpha}-\boldsymbol{e}_d} \dot{\xi}_d n \,\mathrm{d}\boldsymbol{\xi} - \int_{\partial\Omega_{\xi}} \boldsymbol{\xi}^{\boldsymbol{\alpha}} n\dot{\boldsymbol{\xi}}\cdot\mathbf{n}_{\xi} \,\mathrm{d}S_{\xi} + \int_{\Omega_{\xi}} \boldsymbol{\xi}^{\boldsymbol{\alpha}} \mathcal{S} \, \mathrm{d}\boldsymbol{\xi}.
\end{equation}
As outlined in Section~\ref{sec:architecture}, QMOMKit treats the terms on the right-hand side in two different ways. The first term represents the interior contribution of phase-space advection and describes the continuous motion of the population through phase space. This motion through phase space can be evaluated by advancing the abscissas of the quadrature nodes according to $\dot{\boldsymbol{\xi}}$ while keeping their weights unchanged \citep{Yuan2012,Heylmun2021,Li2022c}. The second term is the corresponding flux through the phase-space boundary and the third term represents sources or sinks within the phase space, which are both terms that can change the population itself. Since moving existing abscissas cannot readily represent such population changes, QMOMKit applies these contributions directly to the moments.

These two different numerical treatments motivate a splitting of the local phase-space evolution \citep{Yuan2011, Yuan2012, Nguyen2016, Patel2019a, Heylmun2021}. The interior contribution of phase-space advection is advanced through the motion of the quadrature abscissas in phase space and defines the advection subproblem,
\begin{equation}
\mathcal{B}_{\mathrm{adv},\boldsymbol{\alpha}} = \sum_{\substack{d=1\\\alpha_d>0}}^{D} \alpha_d \int_{\Omega_{\xi}} \boldsymbol{\xi}^{\boldsymbol{\alpha}-\boldsymbol{e}_d} \dot{\xi}_d n \, \mathrm{d}\boldsymbol{\xi},
\label{eq:xi-advection}
\end{equation}
whereas the phase-space boundary flux and the population source are grouped into the moment-source subproblem, 
\begin{equation}
\mathcal{B}_{\mathrm{src},\boldsymbol{\alpha}} = - \int_{\partial\Omega_{\xi}} \boldsymbol{\xi}^{\boldsymbol{\alpha}} n\dot{\boldsymbol{\xi}}\cdot\mathbf{n}_{\xi} \,\mathrm{d}S_{\xi} + \int_{\Omega_{\xi}} \boldsymbol{\xi}^{\boldsymbol{\alpha}} \mathcal{S} \, \mathrm{d}\boldsymbol{\xi}.
\label{eq:mte-sources}
\end{equation}
such that $\mathcal{B}=\mathcal{B}_{\mathrm{adv}}+\mathcal{B}_{\mathrm{src}}$. Let $\Phi_{\mathrm{adv}}^{\Delta t}$ and $\Phi_{\mathrm{src}}^{\Delta t}$ denote the numerical evolution maps obtained by advancing the corresponding subproblems over a time step $\Delta t$. QMOMKit offers the two Lie splittings \citep{Descombes2007}
\begin{equation}
\Phi^{\Delta t} = \Phi_{\mathrm{adv}}^{\Delta t} \circ \Phi_{\mathrm{src}}^{\Delta t},
\label{eq:lie-source-advection}
\end{equation}
and
\begin{equation}
\Phi^{\Delta t} = \Phi_{\mathrm{src}}^{\Delta t} \circ \Phi_{\mathrm{adv}}^{\Delta t},
\label{eq:lie-advection-source}
\end{equation}
as well as the symmetric Strang splitting \citep{Strang1968}
\begin{equation}
\Phi^{\Delta t} = \Phi_{\mathrm{adv}}^{\Delta t/2} \circ \Phi_{\mathrm{src}}^{\Delta t} \circ \Phi_{\mathrm{adv}}^{\Delta t/2}.
\label{eq:strang-advection-source}
\end{equation}
Provided that the subproblems are integrated with sufficient accuracy, the Lie splittings are first-order accurate in time and the Strang splitting, which is the default in QMOMKit, is second-order accurate \citep{Spiteri2026}. Similar splitting strategies are commonly used in quadrature-based moment methods \citep{Yuan2011, Heylmun2021, Li2022c, Kong2017, Li2019a, Nguyen2016, Bouguezzoul2026a}.

For phase-space advection, QMOMKit advances the quadrature abscissas according to the user-supplied phase-space velocity $\dot{\boldsymbol{\xi}}(t,\boldsymbol{\xi}_{\mathbf{i}})$, while keeping the weights and the node count of the quadrature unchanged. The moments are then recomputed from the advected quadrature. Because the weights remain positive, the moments of the advected quadrature correspond to a non-negative distribution and are therefore realizable on the real line \citep{Yuan2012}. QMOMKit provides several Runge-Kutta schemes for integrating the abscissa trajectories: The RK23 scheme of \citet{Bogacki1989} and the RK45 scheme of \citet{Dormand1980} use embedded lower-order estimates to adapt the internal time step, whereas the SSP-RK3 scheme \citep{Shu1988,Gottlieb1998} advances the abscissas with a prescribed time step. By default, the RK23 scheme is used in QMOMKit.

The moment-source subproblem advances the moments directly according to Eq.~\eqref{eq:mte-sources}. Following \citet{Nguyen2016}, QMOMKit integrates the moment-source subproblem using the SSP-RK3 scheme, with optional moment-realizability checks (Section~\ref{sec:moment-realizability}). If a source model requires moments that are not part of the required moment set, such as moments of higher or fractional order, QMOMKit evaluates them using the most recent quadrature. After the source update, the quadrature is reconstructed from the updated moments. Because the updated moment set may admit a different number of nodes, this reconstruction may change the active quadrature topology.

\subsection{Selecting the active node count}
\label{sec:node-count-selection}

In the adaptive mode, which is the default in QMOMKit, the configured node count $N^{\max}$ is an upper limit, whereas the active node count is selected each time a quadrature is reconstructed from new or updated moments. A candidate quadrature $\mathcal{Q}_N$ is accepted if it passes the degeneracy test in Eq.~\eqref{eq:wheeler-degeneracy-test} or Eq.~\eqref{eq:cqmom-degeneracy-test}, as well as the criteria in Eqs.~\eqref{eq:quadrature-acceptance-criteria-w}--\eqref{eq:quadrature-acceptance-criteria-Omega}.

The first reconstruction of a state starts from $N^{\max}$. Whenever a one-dimensional quadrature has a previously accepted active node count $N$, QMOMKit begins the new reconstruction using that node count. If the candidate quadrature $\mathcal{Q}_N$ is accepted and $N<N^{\max}$, QMOMKit also tests a candidate quadrature with one additional node. The active node count increases to $N+1$ only if this candidate quadrature with one additional node is also accepted. If $\mathcal{Q}_N$ is rejected, successively smaller node counts are tested until an acceptable quadrature is found. Hence, the node count $N^{+}$ selected for the new quadrature is
\begin{equation}
N^{+} =
\begin{cases}
N+1, & N<N^{\max},\quad \mathcal{Q}_{N}\ \text{and}\ \mathcal{Q}_{N+1}\ \text{are accepted},
\\[2mm]
N, & \mathcal{Q}_{N}\ \text{is accepted,}\ N<N^{\max},\ \text{and}\ \mathcal{Q}_{N+1}\ \text{is rejected},
\\[2mm]
N, & \mathcal{Q}_{N}\ \text{is accepted and}\ N=N^{\max},
\\[2mm]
\displaystyle \max\left\{N^{\star}\in\mathbb{N}: 1\leq N^{\star}<N,\  \mathcal{Q}_{N^{\star}}\ \text{is accepted} \right\}, & \mathcal{Q}_{N}\ \text{is rejected}.
\end{cases}
\label{eq:node-count-after-source}
\end{equation}
Starting from the previously accepted node count avoids repeatedly testing the configured maximum $N^{\max}$ when fewer nodes are currently admissible. At the same time, limiting growth to one additional node per update allows the quadrature resolution to increase progressively as quadratures with more nodes become acceptable. For CQMOM, the adaptive search is performed independently on each branch of the quadrature hierarchy. Different conditional branches may therefore retain, lose, or gain nodes independently. In a coupled simulation, a new state is created from the transported moments at every time step of the host, as discussed in Section~\ref{sec:interface-case-state}, so that the node-count history applies within each phase-space update but not across time steps of the host.

The applied procedure is deliberately transparent: a rejected candidate quadrature is not altered by clipping out-of-bounds abscissas, merging nearby nodes, or applying other silent corrections. Instead, the candidate quadrature is discarded and a new candidate quadrature is reconstructed with a different node count. As a consequence, the accepted quadrature reflects the explicit adaptive search and acceptance criteria described above.

\subsection{Moment realizability}
\label{sec:moment-realizability}

Moment realizability is concerned with whether a given moment set can arise from a non-negative population distribution on a given support. It is, thus, a property of the moments  and provides a basic admissibility condition for moment-based population models. A successful adaptive reconstruction of a quadrature does not establish realizability, since unrealizable moments might not be considered during quadrature reconstruction with a reduced number of nodes. Realizability may therefore be checked separately.

QMOMKit provides an optional realizability check for truncated one-dimensional moment sets on the real line (Hamburger moment problem), the non-negative real line (Stieltjes moment problem), and bounded intervals (Hausdorff moment problem) \citep{Schmudgen2017}. The check tests whether the matrices built from the moments, the so-called Hankel matrices, are positive semidefinite, as they must be for the moments of a non-negative distribution. If the support has a lower bound $a$ or an upper bound $b$, the same test is applied to matrices built from the shifted moments $M_{\ell+1}-aM_{\ell}$ or $bM_{\ell}-M_{\ell+1}$, respectively. A change of coordinate replaces each of these matrices by a positive multiple of a congruent matrix, which by Sylvester's law of inertia has the same numbers of positive, negative, and zero eigenvalues \citep{Sylvester1852}. The matrices are therefore evaluated in a coordinate that is centred on the mean and scaled by the spread of the distribution, so that the numerical decision depends neither on the location of the distribution nor on the unit of its coordinate. Besides strictly realizable and non-realizable moment sets, the check identifies moment sets on the boundary of the realizable set, such as those of a $k$-point distribution \citep{Schmudgen2017,Putz2023,Laurent2017}, and moment sets for which rounding errors prevent a reliable decision. Boundary moment sets are accepted by default, whereas moment sets that cannot be classified reliably are treated as failures unless the selected policy accepts them, in which case they are left to the quadrature reconstruction. For multivariate distributions, the realizability of the marginal moment set for each dimension may be checked. Satisfaction of these marginal conditions is necessary, but does not in general establish realizability of the full joint moment set.

QMOMKit can apply these realizability checks during the moment-source update of Eq.~\eqref{eq:mte-sources}. Depending on the selected policy, realizability may be checked only after a complete time step of the SSP-RK3 scheme or after every Runge-Kutta stage. Moment realizability after a time step satisfies
\begin{equation}
\mathbf{M}^{n+1}\in\mathcal{R}_{\Omega_{\xi}},
\end{equation}
whereas moments that are realizable at each RK stage satisfy
\begin{equation}
\mathbf{M}^{(s)}\in\mathcal{R}_{\Omega_{\xi}},
\qquad
s=1,2,3,
\end{equation}
with $\mathcal{R}_{\Omega_{\xi}}$ denoting the set of truncated moment sequences realizable on $\Omega_{\xi}$ and $s$ denotes the stage of the SSP-RK3 scheme. If a checked moment state fails the configured realizability requirement, the moment-source integration is stopped, all intermediate moment states are discarded, and the phase-space update returns the moment state at the beginning of the time step. The host application then decides how to proceed, for example by repeating the step with a smaller time step. Optionally, QMOMKit repeats the source integration itself with successively halved time steps \citep{Nguyen2016}. End-of-step checking permits intermediate Runge-Kutta stages to leave the realizable set provided that the final state is realizable, whereas stage-wise checking rejects the integration as soon as the result of a stage becomes unrealizable. 

\section{Software interface}
\label{sec:software-interface}

This section presents the interface through which applications use QMOMKit. Following the division of responsibilities described in Section~\ref{sec:architecture}, the interface is organized around cases and states, which hold the problem definition and the evolving moments and quadrature, functions that reconstruct a quadrature or advance a state in phase space, and callbacks, through which the application supplies its physical models.

\subsection{A minimal example}
\label{sec:interface-minimal-example}

Listing~\ref{lst:qmom-minimal} illustrates the basic workflow for a one-dimensional QMOM problem. It is a reduced form of the example \texttt{qmom\_1d\_minimal} available in the QMOMKit repository, with header includes, diagnostic output, and file output omitted for clarity. The listing contains the main elements of using QMOMKit: a case defining the initial problem, a state containing the evolving moments and quadrature, application-defined callbacks describing the phase-space physics, and advancement in phase space.

\begin{lstlisting}[style=cpp,float=t, caption={Minimal QMOMKit workflow for one-dimensional phase-space evolution.}, label={lst:qmom-minimal}]
int main()
{
    const std::size_t node_count = 3;
    const std::vector<double> initial_moments{1.0, 2.0, 5.025, 14.195, 42.11255, 127.48675};
    qmomkit::QmomPhaseSpaceEvolutionOptions options;

    const qmomkit::QmomCase qmom_case{qmomkit::MomentSet(initial_moments), node_count};
    qmomkit::QmomState state = qmom_case.initial_state();

    const qmomkit::PhaseSpaceVelocity velocity = 
        [](double, const std::vector<double>& coordinates, std::vector<double>& velocities, std::size_t)
    {
        for (std::size_t node = 0; node < coordinates.size(); ++node)
        {
            velocities[node] = 1.0 - coordinates[node];
        }
    };

    const qmomkit::QmomMomentSource source = [](double, const qmomkit::QmomSourceContext& context)
    {
        std::vector<double> derivative(context.size(), 0.0);
        for (std::size_t order = 0; order < context.size(); ++order)
        {
            const double daughter_factor = std::pow(2.0, 1.0 - static_cast<double>(order));
            derivative[order] = 0.5 * (daughter_factor - 1.0) * context.value(order);
        }
        return qmomkit::QmomMomentRate{std::move(derivative)};
    };

    double time = 0.0;
    const double final_time = 1.0;
    const double time_step = 0.01;

    while (time < final_time)
    {
        const double step = std::min(time_step, final_time - time);
        qmomkit::QmomPhaseSpaceEvolutionResult result = 
            qmomkit::advance_qmom_phase_space(state, time, step, velocity, source, options);
        if (!result.ok())
        {
            return 1;
        }
        state = std::move(result.state);
        time = result.final_time;
    }
}
\end{lstlisting}

The example specifies the configured node count and initial raw moments (lines 3--4). The options for the phase-space evolution are initialized with their default values (line 5) and the \texttt{QmomCase} is constructed from the initial moments and node count (line 7). Calling \texttt{initial\_state()} on this case creates the \texttt{QmomState} and reconstructs its quadrature from the initial moments (line 8). The two callbacks (lines 10--28) define the physical model: the velocity callback specifies $\dot{d}=1-d$ for the particle diameter $d$, while the source callback gives the raw-moment rates associated with binary fragmentation. The time-stepping loop (lines 34--45) calls \texttt{advance\_qmom\_phase\_space()} to advance the state over each time step and stops if a step fails. QMOMKit performs the required quadrature reconstruction and phase-space integration internally and returns the result in a \texttt{QmomPhaseSpaceEvolutionResult} object, from which the updated state is obtained. 

\subsection{Cases and states}
\label{sec:interface-case-state}

A QMOMKit simulation begins with a prescribed initial moment set and reconstruction configuration. QMOMKit stores this problem definition in a \textit{case}, while the quantities that evolve in time are stored in a \textit{state}. A state contains the current moment set together with the corresponding quadrature used during phase-space evolution, and the initial state is created by calling \texttt{initial\_state()} on the corresponding case. The resulting workflow is
\begin{equation}
\texttt{Case}
\;\xrightarrow{\;\texttt{initial\_state()}\;}
\texttt{State}
\;\xrightarrow{\;\text{phase-space evolution}\;}
\texttt{State}^{+}.
\label{eq:interface-case-state-workflow}
\end{equation}

\begin{figure}
    \centering
    \includegraphics[width=1\linewidth]{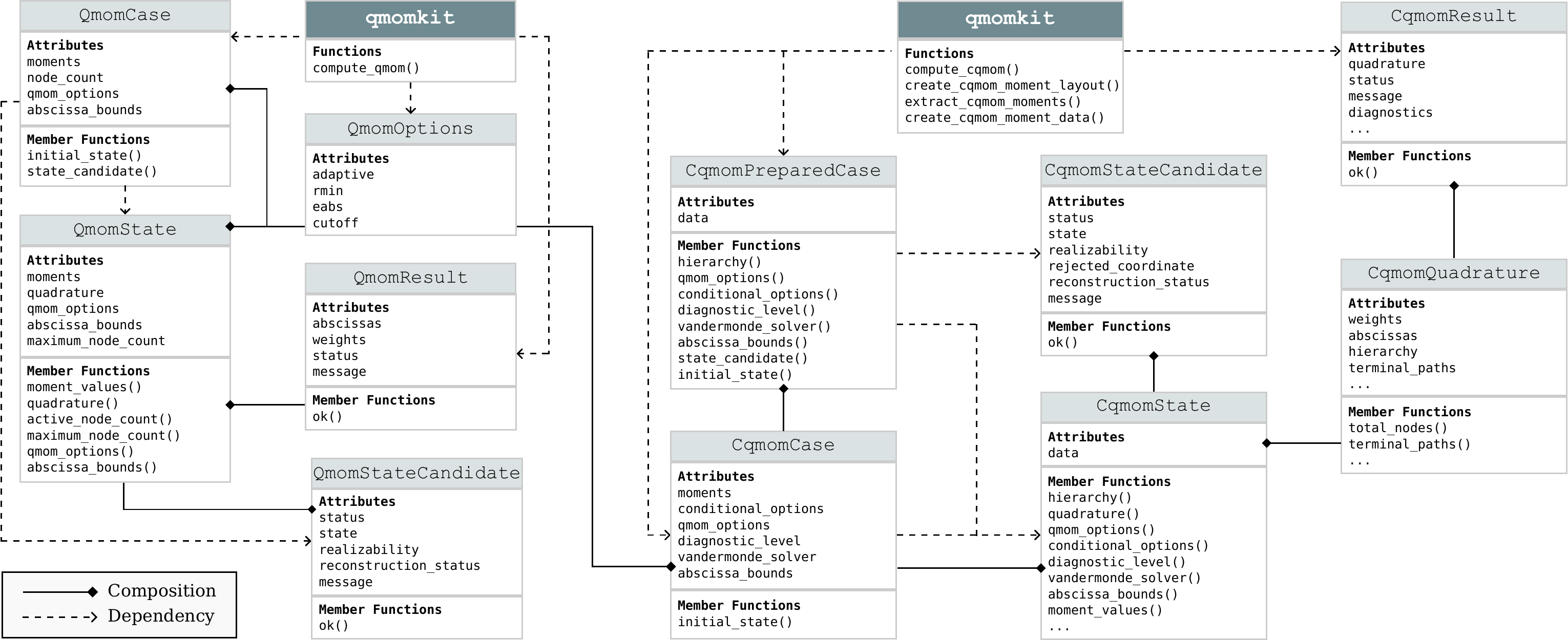}
    \caption{UML diagram of main attributes and member functions for QMOM and CQMOM in QMOMKit.}
    \label{fig:uml-c-qmom}
\end{figure}

For QMOM, a \texttt{QmomCase} contains the initial raw moments, configured maximum node count, and reconstruction options. Calling \texttt{initial\_state()} creates a \texttt{QmomState} containing the current moments and the corresponding quadrature. The state records both the configured maximum and the number of nodes active in the accepted quadrature. Its moments are evaluated from that quadrature so that the two remain consistent and cannot be modified independently.
The same distinction applies to CQMOM. A \texttt{CqmomCase} contains the initial moments, hierarchy, and numerical settings, while a \texttt{CqmomPreparedCase} also holds the recursive branch structure used during reconstruction and can reuse it across successive operations. A \texttt{CqmomState} contains the current multidimensional moments, accepted quadrature, and branch-specific active node counts; the configured node counts remain upper limits. Moment indices and values follow the user-defined phase-space order. In any case, \texttt{initial\_state()} throws an exception if the initial moments cannot be reconstructed, whereas \texttt{state\_candidate()}, described below, reports such a failure as a status.
After initialization, the application passes the current state to the phase-space evolution functions described in Section~\ref{sec:interface-phase-space-evolution}. Figure~\ref{fig:uml-c-qmom} depicts the library architecture for the main QMOM and CQMOM attributes.

A state remains valid as long as only the phase-space evolution changes its moments. In a standalone simulation, such as the example in Listing~\ref{lst:qmom-minimal}, a single state is therefore advanced over the entire simulation. In a coupled simulation, the host changes the moments between two phase-space updates by transporting them in physical space. These moments are the conserved quantities of the spatial discretization and are owned by the host, whereas a state is only required to evolve the moments in phase space. Instead of keeping a state at each location and overwriting its moments, a new state is created at each location from the transported moments $\mathbf{M}_h$ and the settings of the corresponding \textit{case} by calling the function \texttt{state\_candidate()}, which reports the outcome as a status instead of throwing an exception. At each location, the workflow of a coupled simulation is therefore
\begin{equation}
\texttt{State}^{+}
\;\xrightarrow{\;\text{transport by the host}\;}
\mathbf{M}_h
\;\xrightarrow{\;\texttt{state\_candidate()}\;}
\texttt{State}
\;\xrightarrow{\;\text{phase-space evolution}\;}
\texttt{State}^{+}.
\label{eq:interface-host-workflow}
\end{equation}
The lifetime of a state is, thus, limited to one phase-space update, and between two updates the moments are held only by the host. If the new state cannot be created or the phase-space update fails, the host may discard the state, while its own moments remain unchanged. Because an adaptive reconstruction does not reject moments that are not realizable, \texttt{state\_candidate()} allows to first check the realizability of the updated moments, as described in Section~\ref{sec:moment-realizability}. However, keeping the transported moments realizable remains the responsibility of the spatial discretization of the host. A location whose zeroth moment lies below the cutoff defined for the Wheeler algorithm, see Section~\ref{sec:numerics-quadrature}, is reported as \texttt{empty}, such that the host application may decide to forego the phase-space evolution at this location for the next step.

\subsection{Reconstructing a quadrature}

The most direct use of QMOMKit is to reconstruct a quadrature from a prescribed set of raw moments. The functions \texttt{compute\_qmom()} and \texttt{compute\_cqmom()} do so without creating a state or advancing the moments in time. For repeated CQMOM reconstruction, a \texttt{CqmomPreparedCase} avoids rebuilding the recursive branch structure. The returned result contains the reconstructed abscissas and weights together with a status. CQMOM results also describe the conditional branch hierarchy and its multidimensional abscissas of the terminal nodes are returned in the user-defined phase-space order, independently of the conditioning order used during reconstruction. The application should check \texttt{result.ok()} before using the quadrature, since a failed result provides an explanatory message rather than a valid reconstruction.

The direct reconstruction interface is useful when an application requires the quadrature itself but manages the subsequent evolution independently. The returned abscissas and weights can then be used to evaluate integrals over the population distribution within the application's own numerical procedure. In this mode, no \texttt{QmomState} or \texttt{CqmomState} is required, and none of the phase-space evolution functions described below need to be used.

\subsection{Phase-space evolution}
\label{sec:interface-phase-space-evolution}

QMOMKit provides functions for phase-space advection and for moment sources, including phase-space boundary contributions, as well as functions that combines all them. Because these functions can be called individually, a host application can incorporate the phase-space evolution into its own time-integration procedure. Each function can be applied to one \texttt{\{Qmom|Cqmom\}State} or to a batch of independent states.

Phase-space advection is invoked with \texttt{advance\_\{qmom|cqmom\}\_advection()}. The application supplies the current \texttt{QmomState} or \texttt{CqmomState}, the starting time $t$, the step length, the integration options, and a callback defining $\dot{\boldsymbol{\xi}}(t,\boldsymbol{\xi})$. The callback writes the velocity components into an array supplied by QMOMKit. Both the abscissas and velocities follow the user-defined phase-space order. QMOMKit integrates the trajectories of the quadrature abscissas over the step using the RK23, RK45, or SSP-RK3 scheme. During this update, the quadrature weights and active node or branch counts remain unchanged, since only the abscissas evolve. The raw moments are then recomputed from the updated quadrature and stored in the resulting state, ensuring that the moment set and the quadrature  remain mutually consistent.

The moment-source subproblem in Eq.~\eqref{eq:mte-sources} is invoked with \texttt{advance\_\{qmom|cqmom\}\_moment\_source()}. The application supplies the current \texttt{QmomState} or \texttt{CqmomState}, the starting time and step length, the source-integration options, and callbacks defining the moment rates. QMOMKit advances these rates using the SSP-RK3 scheme and reconstructs the quadrature from the updated moments. The callback for the moment source defines a contribution of the form
\begin{equation}
\left. \frac{\mathrm{d}\mathbf{M}}{\mathrm{d}t} \right|_{\mathrm{src}} = \mathbf{{S}}_M,
\end{equation}
where the application supplies the physical model represented by $\mathbf{S}_M$. Such contributions may describe aggregation, breakage, collisions, or other processes that change the population. 
Phase-space boundary contributions are handled by the same moment-space integration. An optional callback for the boundary contribution defines
\begin{equation}
\left. \frac{\mathrm{d}\mathbf{M}}{\mathrm{d}t} \right|_{\mathrm{bnd}} = \mathbf{F}_M,
\end{equation}
so that QMOMKit advances the moment-source subproblem of Eq.~\eqref{eq:mte-sources} in the form
\begin{equation}
\frac{\mathrm{d}\mathbf{M}}{\mathrm{d}t} = \mathbf{{S}}_M + \mathbf{F}_M.
\end{equation}
The phase-space boundary callback represents population entering or leaving the phase-space domain and is distinct from a boundary condition in physical space, which remains the responsibility of the host application.
Optional realizability checks can be applied during the integration of the moment-source subproblem as described in Section~\ref{sec:moment-realizability}.

QMOMKit also provides functions for solving the complete local phase-space evolution. The function \texttt{advance\_\{qmom|cqmom\}\_phase\_space()} accepts the current \texttt{QmomState} or \texttt{CqmomState}, the starting time and step length, the callbacks defining the phase-space velocity, moment source, and boundary contribution, and the numerical options for the corresponding subproblems. After a successful advancement, QMOMKit returns the updated state containing the moments and the corresponding quadrature.
The combined function composes the advection and moment-source updates using one of the operator-splitting schemes introduced in Section~\ref{sec:moment-evolution}. Each substep uses the same numerical procedure as the corresponding standalone advection or moment-source function, while the selected splitting determines the order and duration of the substeps.
On success, the state at $t+\Delta t$ is returned; on failure, the unchanged input state at $t$ is returned. The host can therefore recover without having to reconstruct the previous state.

For host applications that evolve one QMOM or CQMOM state at each host-defined location (e.g., cell center or integration point), batch variants of the phase-space-evolution functions provide a convenient interface to advance a large number of states with a single call. A batch call applies the single-state evolution function serially, in input order, using a common starting time, step length, and set of evolution options. Each state nevertheless retains its own reconstruction settings, adaptive integration and reconstruction behavior, diagnostics, and acceptance status. The ordinary batch functions use the same callbacks for every state, whereas the batch-index-aware variants also pass the position of the state in the batch to each callback. The batch returns one result for each state, allowing the host to check all local updates before accepting the new states or retrying the step. These functions provide a convenient association between states and host data, but do not provide parallel execution or communication.

\subsection{Callbacks}
\label{sec:interface-callbacks}

Callbacks connect QMOMKit's numerical algorithms to application-specific physical models. The application passes the callbacks to the evolution functions; they are not stored in a case or state. Each callback may be a free function, lambda, or function object. Lambdas and function objects may also hold application data.

A phase-space velocity callback receives the coordinates of all quadrature nodes in one array. The values are grouped by node and follow the user-defined order of the phase-space variables. For instance, for a one-dimensional phase space containing only $d$, the array with the coordinates is simply $(d_0,d_1,\ldots)$, whereas the array is $(d_0,u_0,d_1,u_1,\ldots)$ for a two-dimensional phase space $(d,u)$. The callback writes the corresponding components of $\dot{\boldsymbol{\xi}}$ into a preallocated array with the same layout, as illustrated in Listing \ref{lst:qmom-minimal}. The callback must fill every entry without resizing the array.

For moment-source and boundary-contribution callbacks, QMOMKit constructs at each SSP-RK3 stage a source context of type \texttt{\{Qmom|Cqmom\}SourceContext}, representing the moment state being advanced. It gives the callback read-only access to the current moment values and identifies the moment associated with each value. For QMOM, \texttt{QmomSourceContext::value(}$\ell$\texttt{)} is the raw moment $M_\ell$. For CQMOM, \texttt{CqmomSourceContext::index(i)} gives the multi-index of \texttt{CqmomSourceContext::value(i)}. The callback returns the corresponding moment rates in the same order. The stage time is passed as the callback’s first argument and is also available through \texttt{\{Qmom|Cqmom\}SourceContext::time()}. A source model may also request integer- or fractional-order moments that are not stored in the state but can be evaluated from the quadrature reconstructed for that stage. Alternatively, a moment source may traverse the quadrature nodes directly, allowing the source model to use additional moment information. A boundary callback uses the same context and output layout, but its rates represent exchange across the phase-space boundary. When both callbacks are supplied, QMOMKit adds their rates within the same moment-source update.

Callbacks may use data owned by the host application. If, for example, the phase-space velocity callback captures by reference a persistent structure containing the carrier-flow velocity at the two most recent time levels of the host, the callback can interpolate this velocity to each stage time and use it to calculate the velocity of every quadrature node. The callback therefore need not be rebuilt when the host data change, but the referenced object must remain valid for every invocation. QMOMKit may call a callback repeatedly, including when adaptive substeps are rejected or source updates are retried. A callback must therefore not assume one invocation per outer time step or perform irreversible updates on each call.

Listing~\ref{lst:callback-construction} illustrates the construction patterns of the callbacks for a two-dimensional phase space, assuming that the phase-space evolution is handled using the batch functions. The immutable model parameter is captured by value, whereas host-owned fields are captured by reference so that updated values are visible without reconstructing the callbacks. The phase-space velocity callback evaluates the local carrier velocity at the supplied time and writes both components for every quadrature node. The moment-source callback evaluates the source from \texttt{context.index(i)} and \texttt{context.value(i)} and returns one rate per entry. The corresponding single-state callbacks omit the leading batch index. A boundary-contribution callback is constructed in the same way as the source callback and returns the same rate type.

The batch-index-aware callback types add a leading argument, the zero-based position of the state in the batch (\texttt{std::size\_t batch} in Listing~\ref{lst:callback-construction}). In the two-dimensional batch example, the host stores state $i$ and its local field data at position $i$ in their respective arrays, then uses the supplied index to select both. QMOMKit assigns this index no spatial meaning. Every callback output must have the required size and contain only finite values. Otherwise, the evolution fails and the input state is returned unchanged.

\begin{lstlisting}[style=cpp, float=t, caption={Construction of batch-index-aware callbacks for a two-dimensional model with coordinates $(d,u)$.}, label={lst:callback-construction}]
HostData host = make_host_data(); // Outlives both callbacks.
const double diameter_relaxation = 1.0;

const qmomkit::BatchPhaseSpaceVelocity velocity =
    [diameter_relaxation, &host](std::size_t batch, double time, const std::vector<double>& x, 
        std::vector<double>& v, std::size_t dimensions)
{
    const double carrier = carrier_velocity_at(host, batch, time);
    for (std::size_t k = 0; k < x.size(); k += dimensions)
    {
        v[k] = diameter_relaxation * (1.0 - x[k]);
        v[k + 1] = carrier - x[k + 1];
    }
};

const qmomkit::BatchCqmomMomentSource source =
    [&host](std::size_t batch, double, const qmomkit::CqmomSourceContext& context)
{
    const double fragmentation_rate = host.fragmentation_rate[batch];
    std::vector<double> rates(context.size(), 0.0);

    for (std::size_t i = 0; i < context.size(); ++i)
    {
        const double diameter_order = static_cast<double>(context.index(i)[0]);
        const double daughter_factor = std::pow(2.0, 1.0 - diameter_order);
        
        rates[i] = fragmentation_rate *(daughter_factor - 1.0) * context.value(i);
    }

    return qmomkit::CqmomMomentRate{std::move(rates)};
};
\end{lstlisting}

\subsection{Configuration}
\label{sec:interface-configuration}

QMOMKit can be configured either from a \texttt{*.qmom} file or directly through its C++ interface. Loading a configuration reports success or failure, which an application may check before using the configured case.
Reconstruction, phase-space evolution, and output settings are loaded separately even when they appear in the same file. The evolution settings control node advection, realizability checks, and operator splitting.

A configuration file may provide explicit initial moments. Alternatively, QMOMKit may generate univariate NDFs from an analytical distribution or a mixture of distributions; for multivariate NDFs, it may generate the required moments from an independent product of marginal distributions or mixtures. The reconstruction configuration defines the phase-space variables and their order, the configured maximum node count of each coordinate, prescribed supports, and whether the reconstruction is adaptive, which is the default in QMOMKit. For CQMOM, the configuration additionally specifies the conditioning order and the solver for the conditional systems. The weight and separation thresholds of the quadrature reconstruction, see Section~\ref{sec:numerics-quadrature}, can be set through the C++ interface.

\section{Verification of numerical algorithms}
\label{sec:verification}

Following the testing philosophy described in Section~\ref{sec:architecture}, this section examines representative verification studies targeting key numerical components of QMOMKit: one-dimensional and recursive multidimensional quadrature reconstruction, adaptive quadrature topology, and phase-space time integration. Each study compares the numerical results with an analytical or manufactured reference solution, allowing the accuracy and expected behavior of the corresponding algorithm to be assessed quantitatively.

\subsection{One-dimensional quadrature reconstruction}
\label{sec:verification-quadrature}

In QMOMKit, the modified Wheeler algorithm described in Section \ref{sec:numerics-quadrature} is used to reconstruct quadratures. This reconstruction is verified here using three Gaussian quadrature families with different supports and moment growth: Gauss-Legendre for the unit weight on $[-1,1]$, Gauss-Laguerre with $\alpha=0$ and measure $\exp(-\xi)\,\mathrm{d}\xi$ on $[0,\infty)$, and Gauss-Hermite for the standard-normal measure $(2\pi)^{-1/2}\exp(-\xi^2/2)\,\mathrm{d}\xi$ on $\mathbb{R}$, using the probabilists' Hermite convention. Their analytical raw moments are
\begin{equation}
M_\ell =
\begin{cases}
2/(\ell+1), & \text{Gauss-Legendre, } \ell \text{ even},\\
0,       & \text{Gauss-Legendre, } \ell \text{ odd},\\
\ell!,      & \text{Gauss-Laguerre},\\
(\ell-1)!!,& \text{Gauss-Hermite, } \ell \text{ even},\\
0,       & \text{Gauss-Hermite, } \ell \text{ odd}.
\end{cases}
\label{eq:verification-analytical-moments}
\end{equation}
These three families provide complementary tests: Gauss-Legendre has bounded support and moderate moments, whereas the Gauss-Laguerre and Gauss-Hermite moments grow rapidly with increasing order. 

Using the Golub-Welsch algorithm \citep{Golub1969}, the reference quadratures with abscissas $\xi_i$ and weights $w_i$ are computed independently in long-double precision and then rounded to double precision to provide the input moments $M_0,\ldots,M_{2N-1}$, an approach adapted from \citet{Putz2023}. QMOMKit reconstructs the quadrature from these moments for each node count $N=1,\ldots,10$ in the non-adaptive mode, so that all $N$ nodes are compared. The reported errors based on the reconstructed pairs $(\widehat{\xi}_i,\widehat{w}_i)$ and the reference pairs $(\xi_i,w_i)$ are
\begin{align}
E_\xi &= \max_i \frac{|\widehat{\xi}_i-\xi_i|}{\max_j|\xi_j|}, \label{eq:verification-abscissa-error} \\
E_w &= \max_i \frac{|\widehat{w}_i-w_i|}{M_0}, \\
E_M &= \max_{0\leq \ell<2N} \frac{|\widehat{M}_\ell-M_\ell|}{\bar{M}_\ell}, \label{eq:verification-moment-error}
\end{align}
where
\begin{equation}
 \widehat{M}_\ell =\sum_{i=1}^{N}\widehat{w}_i\widehat{\xi}_i^\ell.
\end{equation}
The abscissa error is scaled by the largest absolute abscissa of the reference quadrature, the weight error by its total weight, and the moment error by
\begin{equation}
    \bar{M}_\ell=\sum_{i=1}^{N}w_i|\xi_i|^\ell,
\end{equation}
the magnitude of the terms that form the moment. Like the reconstruction itself, these errors do not depend on the unit of $\xi$ or on the number density. Scaling the moment error by $\bar{M}_\ell$ rather than by $|M_\ell|$ matters for the odd moments of the symmetric rules, which vanish, so that a relative error would measure the cancellation between large terms rather than the accuracy of the quadrature. For $N=1$, the single abscissa of the symmetric rules is zero, and the absolute errors are reported where the scale vanishes. The moment comparison covers the same $2N$ moments used to reconstruct the quadrature.

\begin{figure}[t]
\centering
\includegraphics[width=\linewidth] {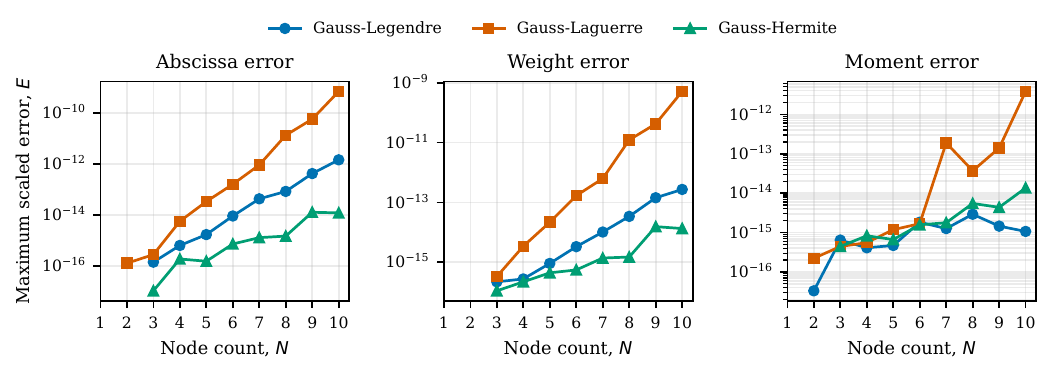}
\caption{Maximum scaled abscissa, weight, and moment errors for the three quadrature families. Each point represents an $N$-node reconstruction from moments of orders $0$ through $2N-1$. With the scaling of Eqs.~\eqref{eq:verification-abscissa-error}--\eqref{eq:verification-moment-error}, the errors do not depend on the unit of the coordinate or on the number density. Exact zero errors are not visible on the logarithmic axes.}
\label{fig:verification-quadrature}
\end{figure}

As observed in Figure \ref{fig:verification-quadrature}, the abscissa and weight errors remain small throughout the tested range, although they generally increase with $N$ as higher-order raw moments make the inversion more sensitive to floating-point perturbations, consistent with the ill-conditioning of the moment inversion previously reported for larger node counts \citep{Yuan2012,Fox2023a,Heylmun2021}. This effect is most pronounced for the Gauss-Laguerre quadrature, whose moments grow with $\ell!$, with $E_\xi=6.9\times10^{-10}$ and $E_w=5.2\times10^{-10}$ for $N=10$, and weakest for the Gauss-Hermite quadrature. The moment errors are considerably smaller: the reconstructed quadratures reproduce their input moments with $E_M<4\times10^{-12}$ for all families and node counts. This difference is consistent with the conditioning of the moment inversion, in which small changes of the moments cause larger changes of the abscissas and weights.

\subsection{Recursive CQMOM reconstruction}
\label{sec:verification-recursive-cqmom}

The dimension-independent recursive CQMOM implementation described in Section \ref{sec:numerics-cqmom} is verified using a manufactured affine chain with a known hierarchical quadrature. For each phase-space dimension $d$, let $z_d$ be an independent discrete variable with value $-5/8$, $1/8$, $3/4$ and corresponding probabilities $1/4$, $1/2$, $1/4$. The phase-space abscissas are defined recursively as
\begin{align}
\xi_1 &= z_1, \\ \xi_d &= \frac{1}{16} + \frac{3}{16}\xi_{d-1} + \frac{13}{16}z_d, \qquad d=2,\ldots,D.
\label{eq:verification-affine-chain}
\end{align}
Each abscissa therefore depends on the preceding abscissa as well as on a new independent variable. For a fixed parent value $\xi_{d-1}$, the conditional distribution of $\xi_d$ contains three known nodes. Repeating this construction produces a correlated, tree-like quadrature with three child nodes on every branch and $3^D$ terminal nodes. The identity conditioning order $\boldsymbol{\pi}=(1,\ldots,D)$ is used and the reconstruction runs in the non-adaptive mode. Reference moments are evaluated independently in long-double arithmetic by recursively expanding Eq.~\eqref{eq:verification-affine-chain} and every terminal node of the reconstruction is compared with the exact weighted point of the same branch. The recursive reconstruction algorithm described in Section \ref{sec:numerics-cqmom} is used in every tested dimension; no dimension-specific procedure is introduced.

The errors are defined with terminal abscissas $\xi_{\mathbf{i},d}$ and weights $W_{\mathbf{i}}$, similar to Section~\ref{sec:verification-quadrature}, as
\begin{align}
E_\xi &= \max_{\mathbf{i},d} \frac{|\widehat{\xi}_{\mathbf{i},d}-\xi_{\mathbf{i},d}|}{\max_{\mathbf{j}}|\xi_{\mathbf{j},d}|}, \label{eq:cqmom-verification-abscissa-error} \\
E_w &= \max_{\mathbf{i}} \frac{|\widehat{W}_{\mathbf{i}}-W_{\mathbf{i}}|}{M_{\boldsymbol{0}}}, \\
E_M &= \max_{\boldsymbol{\alpha}} \frac{|\widehat{M}_{\boldsymbol{\alpha}}-M_{\boldsymbol{\alpha}}|}{\bar{M}_{\boldsymbol{\alpha}}},
\label{eq:cqmom-verification-moment-error}
\end{align}
where the abscissa error of each coordinate is scaled by the largest absolute abscissa of that coordinate, the moment error is maximized over the required moment set, and $\bar{M}_{\boldsymbol{\alpha}}=\sum_{\mathbf{i}\in\mathcal{T}_D}W_{\mathbf{i}}\prod_{d=1}^{D}|\xi_{\mathbf{i},d}|^{\alpha_d}$. Table~\ref{tab:verification-cqmom-affine} reports these errors. The manufactured affine chain is reconstructed accurately through $D=10$: the weight errors remain at the level of rounding errors, the moment errors stay below $2\times10^{-13}$, and the abscissa errors reach $1.6\times10^{-12}$. The growth of these rounding errors with $D$ reflects the increasing number of recursive conditional solves through which they propagate. 

\begin{table}[t]
\centering
\caption{Recursive CQMOM quadrature reconstruction of the manufactured affine chain. Each coordinate uses three nodes. The errors of the terminal abscissas, $E_\xi$, of the terminal weights, $E_w$, and of the moments, $E_M$, are defined in Eqs.~\eqref{eq:cqmom-verification-abscissa-error}--\eqref{eq:cqmom-verification-moment-error}.}
\label{tab:verification-cqmom-affine}
\begin{tabular}{rrrccc}
\hline
$D$ & Terminal nodes & Required moments & $E_\xi$ & $E_w$ & $E_M$ \\
\hline
2  & 9     & 21      & $4.44\times10^{-16}$ & $1.39\times10^{-16}$ & $8.09\times10^{-16}$ \\
3  & 27    & 66      & $3.50\times10^{-15}$ & $5.27\times10^{-16}$ & $9.85\times10^{-16}$ \\
5  & 243   & 606     & $2.32\times10^{-14}$ & $2.88\times10^{-16}$ & $1.25\times10^{-15}$ \\
7  & 2187  & 5466    & $1.19\times10^{-13}$ & $9.01\times10^{-16}$ & $6.67\times10^{-15}$ \\
10 & 59049 & 147621  & $1.56\times10^{-12}$ & $1.25\times10^{-15}$ & $1.76\times10^{-13}$ \\
\hline
\end{tabular}
\end{table}

\subsection{Adaptive branch-specific quadrature}
\label{sec:verification-adaptive-branches}

The adaptive branch-specific quadrature described in Section \ref{sec:numerics-cqmom} is examined with a test that forces QMOMKit to adapt the active node count when a conditional distribution first contracts, is subsequently broadened by a moment source, and then contracts again. The two phase-space coordinates are assumed to be particle size $d$ and velocity $u$. The size coordinate is fixed at $d=d_0=2$ and is represented by a single marginal node. Conditional on this node, the initial velocity moments are those of $u\sim\mathcal{N}(1,1)$. The configured maximum node counts are one for $d$ and five for $u$, with conditioning order $(d,u)$.

The phase-space velocities are $\dot{d} = 0$ and $\dot{u} = -5(u-u_\mathrm{eq})$, with equilibrium velocity $u_\mathrm{eq}=0.5$, such that the size distribution remains unchanged, whereas the velocity mean and variance satisfy $\overline{u}(t) = u_\mathrm{eq} + (1-u_\mathrm{eq})\exp(-5t)$ and $\operatorname{Var}[u](t) = \exp(-10t)$, respectively. As the velocity variance approaches zero while the mean velocity approaches $u_\mathrm{eq}$, the represented distribution approaches a single velocity value. The adaptive reconstruction should therefore replace the initial five-node velocity quadrature with progressively fewer active nodes, eventually reaching a single node. A nonzero equilibrium velocity is chosen deliberately: relaxation towards $u=0$ would merely rescale every velocity by the same factor $\exp(-5t)$, which is equivalent to a change of the velocity unit rather than a contraction of the distribution relative to its mean. A moment source is then applied over the time interval $t\in[4,4.05)$, representing a symmetric velocity-splitting event: At rate $\lambda=20$, the contribution located at $(d,u)$ is replaced by two equal contributions at $(d,u-\Delta u)$ and $(d,u+\Delta u)$, with $\Delta u=0.10$. For the mixed moment $M_{p,q}$, the resulting source is
\begin{equation}
\left. \frac{\mathrm{d}M_{p,q}}{\mathrm{d}t} \right|_{\mathrm{src}} = \lambda \sum_{\substack{r=2\\ r\ \mathrm{even}}}^{q} \binom{q}{r} (\Delta u)^r M_{p,q-r}. \label{eq:verification-moment2}
\end{equation}
The sum is empty for $q=0$ and $q=1$, so the source leaves the total weight, all velocity-independent moments, and the mean velocity unchanged. For $q=2$, Eq.~\eqref{eq:verification-moment2} yields
\begin{equation}
\left. \frac{\mathrm{d}M_{p,2}}{\mathrm{d}t} \right|_{\mathrm{src}} = \lambda(\Delta u)^2M_{p,0},
\end{equation}
and, hence, introduces positive velocity variance without changing the mean. This broadening of the underlying distribution allows the conditional velocity quadrature to increase the velocity node count.

\begin{figure}[t]
\centering
\includegraphics[width=0.75\linewidth]
{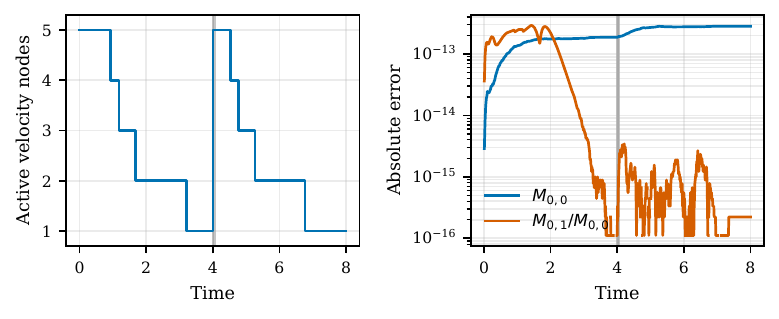}
\caption{Evolution of the active velocity-node count and the corresponding moment errors. The conditional velocity quadrature initially contains five nodes and reduces to one as the velocity variance decays. The shaded interval (gray) marks the symmetric moment source, which introduces velocity variance and allows the velocity quadrature to use more nodes. After the source ends, the continuing velocity relaxation again reduces the active count.}
\label{fig:verification-adaptive}
\end{figure}

Figure~\ref{fig:verification-adaptive} shows the evolution of the quadrature node count and the reconstruction error of the number density $M_{0,0}$ and the mean velocity $M_{0,1}/M_{0,0}$. The conditional velocity quadrature begins with five active nodes and reduces to one as the velocity variance decays. During the injection of velocity variance, the active count increases back to five. Once the source is removed, the same relaxation again drives the velocity quadrature towards a single node. Throughout this process, the reconstructed number density and the mean velocity remain close to their analytical values.

\subsection{Operator-splitting convergence}
\label{sec:verification-splitting}

Here we verify the temporal convergence order of the Lie and Strang splitting schemes used to combine phase-space advection and moment sources, as described in Section \ref{sec:moment-evolution}. Dedicated verification tests of QMOMKit confirm that the convergence rates of the implemented RK schemes is third order for the SSP-RK3 and RK23 schemes, and fifth order for the RK45 scheme, corresponding to the expected orders of these schemes \citep{Shu1988, Bogacki1989, Dormand1980}.

The initial distribution consists of two weighted points with weights $0.45$ and $0.55$ at $\xi=-0.6$ and $\xi=1.4$, whose raw moments are $(M_0,M_1,M_2,M_3) = ( 1,0.5,1.24,1.412)$.
The quadrature nodes move with the constant phase-space velocity $\dot{\xi}=0.45$, for which the resulting contribution to the moments is
\begin{equation}
\left[\mathcal{B}_{\mathrm{adv}}(\mathbf{M}) \right]_\ell = 0.45\,\ell M_{\ell-1}, \qquad \ell=1,\ldots,3,
\end{equation}
with $\left[ \mathcal{B}_{\mathrm{adv}}(\mathbf{M}) \right]_0 = 0$. The moment-source operator is
\begin{equation}
\left[\mathcal{B}_{\mathrm{src}}(\mathbf{M})\right]_\ell= -0.17\,\ell M_\ell, \qquad \ell=0,\ldots,3,
\end{equation}
such that the unsplit moment equations are
\begin{equation}
\frac{\mathrm{d}M_\ell}{\mathrm{d}t} = 0.45\,\ell M_{\ell-1} - 0.17\,\ell M_\ell, \qquad \ell=1,\ldots,3,
\label{eq:splitting-verification-problem}
\end{equation}
together with $\mathrm{d}M_0/\mathrm{d}t=0$. The two operators do not commute: Advection transfers information from $M_{\ell-1}$ to $M_\ell$, whereas the source attenuates each moment at a rate proportional to its order. Applying the operators in a different order over the same time interval therefore gives a different result. The unsplit problem describes the two points moving with $\dot{\xi}=0.45-0.17\,\xi$ at constant weights, so that its exact solution is again a two-point distribution, which a two-node quadrature represents exactly. 

\begin{figure}[t]
\centering
\includegraphics[width=0.5\linewidth]
{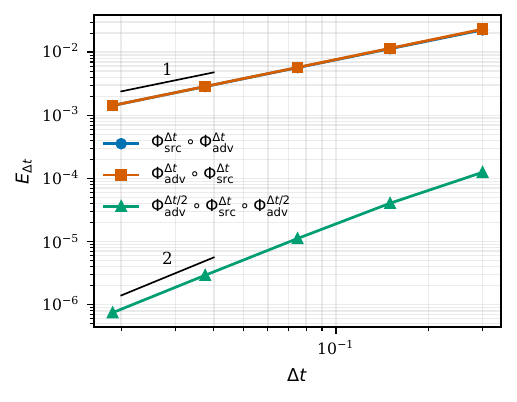}
\caption{Convergence of the operator-splitting schemes available in QMOMKit for the manufactured moment problem in Eq.~\eqref{eq:splitting-verification-problem}. The two Lie splittings and the Strang splitting converge with their expected orders. Note that the results of the two Lie splittings are almost identical and, hence, their lines nearly coincide with each other.}
\label{fig:verification-splitting}
\end{figure}

Let $\Phi_{\mathrm{adv}}^{\Delta t}$ denote the numerical solution of the advection subproblem over an interval of duration $\Delta t$ and let $\Phi_{\mathrm{src}}^{\Delta t}$ denote the corresponding solution of the moment-source subproblem, as previously introduced in Section \ref{sec:moment-evolution}. Using the three operator splittings available in QMOMKit, the advection subproblem is integrated with the default RK23 scheme, which moves the nodes exactly because the phase-space velocity is constant, whereas the source subproblem is integrated with the SSP-RK3 scheme. The considered time steps are $\Delta t \in \left\{ 0.3,\, 0.15,\, 0.075,\, 0.0375,\, 0.01875 \right\}$,
and the final time is $T=1.2$. As in Section~\ref{sec:verification-quadrature}, the error is measured as the maximum difference between the numerical and exact raw moments, scaled by the magnitude of the terms that form each exact moment:
\begin{align}
E_{\Delta t} &= \max_{0\leq \ell\leq 3} \frac{ \left| \widehat{M}_{\ell,\Delta t}(T)-M_\ell(T) \right| }{ \bar{M}_\ell(T) },
\\
\bar{M}_\ell(T) &= \sum_{i=1}^{2}w_i\,|\xi_i(T)|^\ell,
\end{align}
where $\xi_i(T)$ are the positions of the two exact points at time $T$.
Figure~\ref{fig:verification-splitting} shows the convergence of the three operator splittings for a decreasing time step. Both Lie splittings converge with an order of $1.00$, matching their expected first-order convergence, whereas the Strang splitting is observed to converge with an order of $1.97$, approaching the expected second-order convergence. In addition, Strang splitting results in a substantially reduced absolute error compared to both Lie splittings, two to three orders of magnitude smaller in the considered case.

\section{Spatially homogeneous application examples}
\label{sec:applications}

The spatially homogeneous examples presented in this section demonstrate the capabilities and usage of QMOMKit on population models of increasing phase-space dimension and complexity. 
In each example, the QMOMKit solution is compared with a reference solution computed using the Monte Carlo companion code described in \ref{sec:monte-carlo}, obtained by directly evolving a finite sample of the same prescribed population. These comparisons provide a numerical validation of the quadrature reconstruction and phase-space evolution, rather than a physical validation of the underlying application models.

All examples follow the same initialization and comparison procedure. A sample of $10{,}000$ realizations is drawn from the prescribed initial distributions. The corresponding raw moments initialize the QMOMKit simulation, while the same sample is evolved directly with the Monte Carlo companion. Phase-space advection is integrated with the RK23 scheme in both QMOMKit and Monte Carlo simulations. Each example is evolved from $t=0$ to its specified final time $T$ using 100 equal outer time steps, $\Delta t = 0.01T$.
The RK23 integrator may take smaller internal time steps within each outer time step to satisfy its error tolerances. For examples that also contain moment-source or phase-space boundary contributions, $\Delta t$ is the time step over which the selected operator splitting is applied.

\subsection{Evaporation of respiratory aerosols}
\label{sec:applications-aerosol}

\begin{figure}[t]
\centering
\includegraphics[width=0.85\linewidth]
{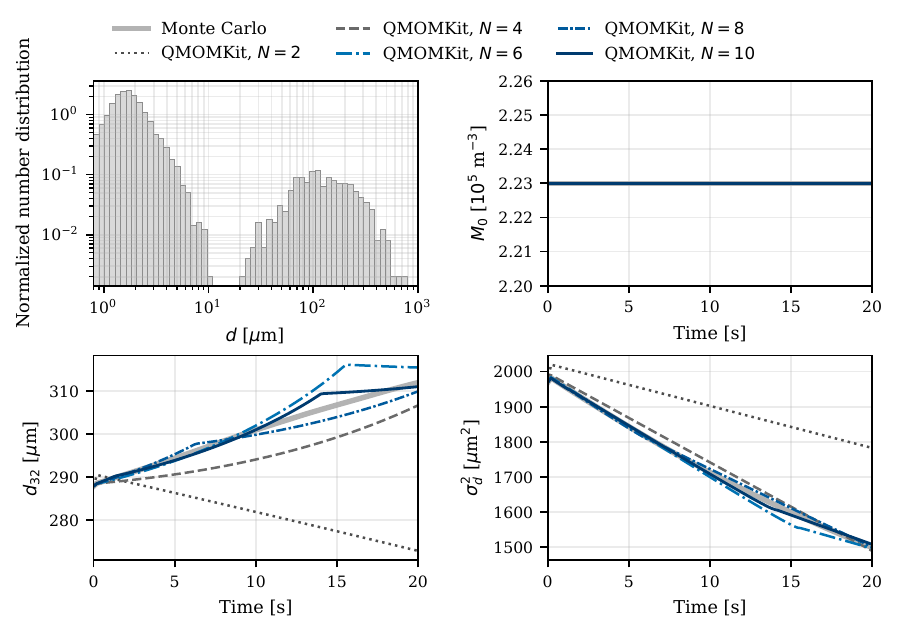}
\caption{Evaporation of a population of respiratory aerosols. The logarithmic histogram shows the initial diameter distribution used by both QMOMKit and the Monte Carlo reference. The number density $M_0$, the Sauter mean diameter $d_{32}=M_3/M_2$, and the diameter variance $\sigma_d^2$ are compared between the Monte Carlo reference and fixed QMOM quadratures with $N=2,4,6,8,$ and $10$ nodes. Droplets reaching $d_{\min}=0.8$~$\mu\mathrm{m}$ remain at this residual diameter.}
\label{fig:applications-aerosol}
\end{figure}

The evaporation of respiratory aerosols provides an illustrative application of QMOM, with the droplet diameter $d$ as the phase-space coordinate. The initial population follows the trimodal bronchiolar-laryngeal-oral (B.L.O.) distribution measured for voluntary coughing \citep{Johnson2011}. The three log-normal modes have median diameters of $1.6$, $1.7$, and $123$~$\mu\mathrm{m}$, respectively, and span a broad range from sub-micrometer aerosols to large respiratory droplets. A diameter-based QMOM evolution with a finite residual size of aerosol droplets has previously been used in population-balance models of such respiratory droplets \citep{Feng2023}. 
Here, the droplet diameter follows the $d^2$ evaporation law \citep{Langmuir1918}
\begin{equation}
\frac{\mathrm{d}d}{\mathrm{d}t} =
\begin{cases}
-K/(2d), & d>d_{\min},\\
0, & d\leq d_{\min},
\end{cases}
\label{eq:applications-aerosol-evaporation}
\end{equation}
with $K=500\,\mu\mathrm{m}^2/\mathrm{s}$ and $d_{\min}=0.8\,\mu\mathrm{m}$, a meaningful representation of the evaporation dynamics of respiratory droplets \citep{DallaBarba2021,Feng2023}. Droplets reaching $d_{\min}$, representing the size of the remaining non-volatile biological material \citep{Feng2023}, do not further reduce in size. A sample of $10{,}000$ droplets is drawn from the B.L.O. distribution and samples outside the modeled interval $0.8\leq d\leq1000~\mu\mathrm{m}$ are discarded, leaving $9{,}575$ droplets for the QMOMKit and Monte Carlo simulations presented below. QMOM quadratures with a fixed node count of $N=2,4,6,8,$ and $10$ are considered, allowing the influence of the quadrature resolution to be assessed. 

Figure~\ref{fig:applications-aerosol} compares the resulting number density, Sauter mean diameter $d_{32}=M_3/M_2$, and diameter variance $\sigma_d^2 = M_2/M_0-(M_1/M_0)^2$. The number density $M_0$ is preserved in all cases, since evaporation changes the size of the droplets, but all droplets retain a finite size and stay part of the distribution. Even though all droplets shrink or remain at $d_{\min}$, the Sauter mean diameter increases. This can be explained by observing that most droplets belong to the much smaller B and L modes, and rapidly accumulate at the residual diameter, where their contributions to $M_2$ and  $M_3$ become small. The larger O-mode droplets therefore become increasingly dominant in the ratio $M_3/M_2$ and the diameter variance decreases as the population accumulates at $d_{\min}$; by $T=20$~s, approximately $96\%$ of the droplets have reached $d_{\min}$.

The two-node quadrature is evidently insufficient to represent the broad trimodal population, whereas four nodes already capture much of its evolution. The higher-resolution quadratures remain close to the Monte Carlo reference, although the error does not decrease monotonically with node count, consistent with the observations of \citet{Grosch2007} for processes that act locally on the particle size. The visible changes in slope of some QMOM curves arise when individual quadrature nodes reach $d_{\min}$ and cease to move. A finite quadrature contains only a small number of such arrival times, whereas the Monte Carlo ensemble contains thousands and, hence, produces a smoother aggregate response. The example thus illustrates both the effectiveness and the characteristic discrete behavior of a fixed QMOM representation for a broad multimodal population approaching a phase-space boundary.

\subsection{Cavitation bubbles}
\label{sec:applications-cavitation}

\begin{figure}[t]
\centering
\includegraphics[width=\linewidth]{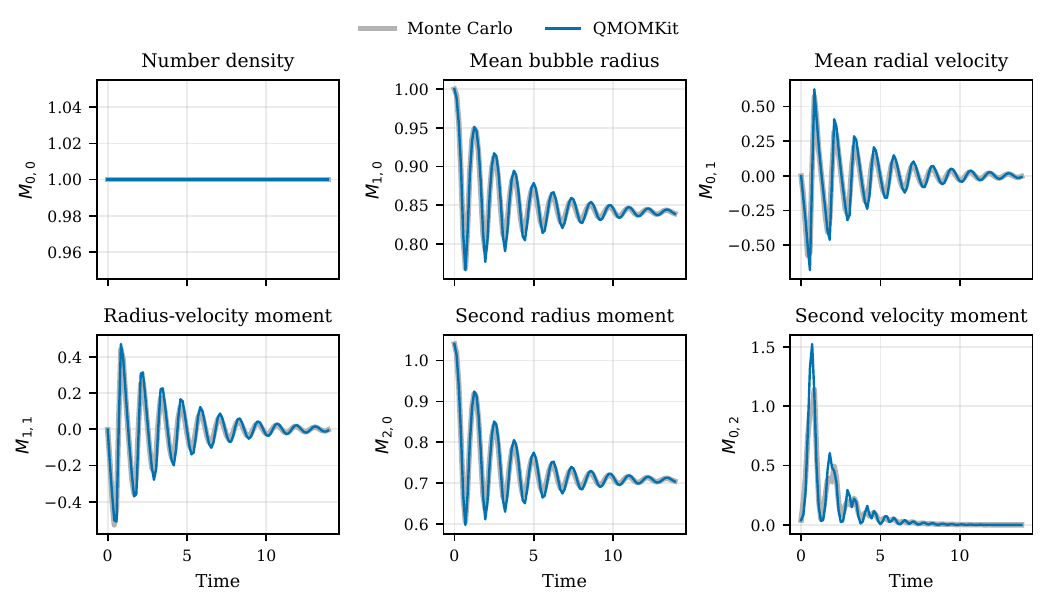}
\caption{Evolution of selected moments for a population of cavitation bubbles, compared with the Monte Carlo reference. QMOMKit reproduces the damped oscillatory dynamics and asymptotic behavior of the first- and second-order moments over the full simulation interval.}
\label{fig:applications-cavitation}
\end{figure}

Inspired by the work of \citet{Bryngelson2023}, we consider a population of cavitation bubbles. The bubble population is represented in the dimensionless phase space $\boldsymbol{\xi}=(R,U)$, where $R$ is the bubble radius and $U=\mathrm{d}R/\mathrm{d}t$ is the radial velocity of the bubble wall. The dynamic behavior of the bubbles is described by a dimensionless Rayleigh-Plesset model \citep{Bryngelson2023},
\begin{align}
\dot{R} &= U, \\
\dot{U} &= \frac{1}{R}\left[\left(\frac{R_0}{R}\right)^{3\gamma}-\frac{1}{C_p}
-\frac{2}{\mathrm{We}R_0}\left(\frac{R_0}{R}-\left(\frac{R_0}{R}\right)^{3\gamma}\right)
-\frac{3}{2}U^2-\frac{4}{\mathrm{Re}}\frac{U}{R}\right],
\label{eq:applications-cavitation}
\end{align}
where the bubbles are assumed to contain a polytropic ideal gas, $R_0=1$ is the equilibrium radius of all bubbles, the driving pressure coefficient is $C_p=0.3$, the polytropic exponent of the gas is $\gamma=1.4$, and the Reynolds and Weber numbers are $\mathrm{Re}=10$ and $\mathrm{We}=1.39$, respectively. The initial population has unit number density, the initial radius $R$ follows a log-normal distribution with $\mu_{\log R}=-0.02$ and $\sigma_{\log R}=0.2$, and the initial radial velocity $U$ follows a normal distribution with zero mean and standard deviation $0.2$. CQMOM is configured with two nodes for each phase-space dimension and conditioning order $(R,U)$. 

As observed in Figure~\ref{fig:applications-cavitation}, QMOMKit reproduces the Monte Carlo reference well throughout the simulated interval. The number density $M_{0,0}$ remains constant, as expected in the absence of population sources. The mean radius $M_{1,0}$ and mean radial velocity $M_{0,1}$ undergo damped oscillations, with the oscillations of all moments correctly decaying towards asymptotic values. Both the frequency and damping of the transient dynamics are captured faithfully.

\subsection{Nucleating and condensing water droplets}
\label{sec:applications-droplets}

\begin{figure}[tp]
\centering
\includegraphics[width=\linewidth]
{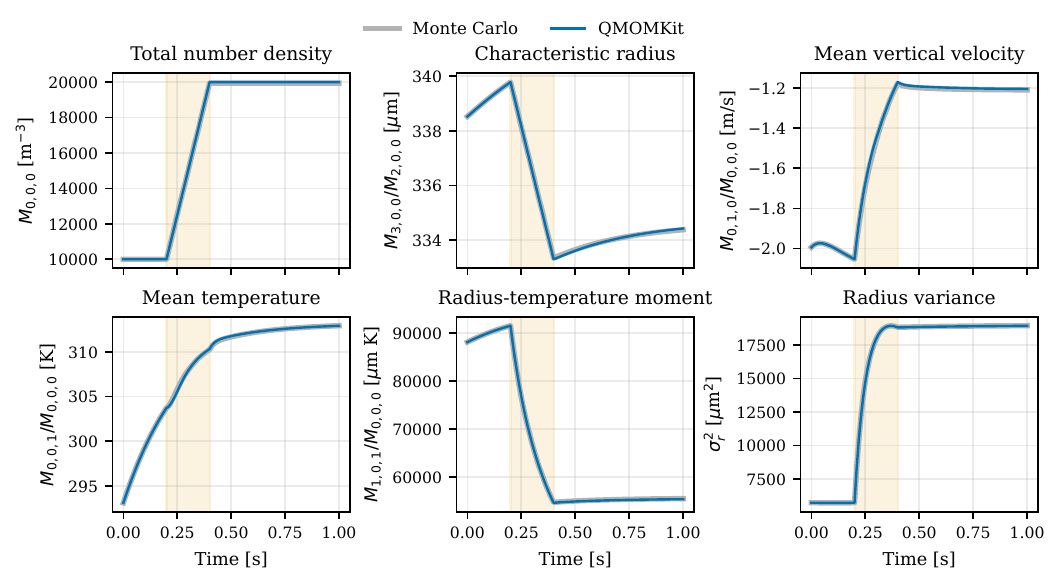}
\caption{Evolution of selected moments and moment-derived quantities for condensing droplets in the three-dimensional phase space $(r,w,T)$, compared with the Monte Carlo reference. New droplets nucleate between $0.20$ and $0.40$~s, indicated by the shaded region, doubling the total population.}
\label{fig:applications-droplets}
\end{figure}

This example considers a population of falling water droplets described by the three-dimensional phase space $\boldsymbol{\xi}=(r,w,T)$, where $r$ is the droplet radius, $w$ is the vertical velocity of the droplet, and $T$ is the droplet temperature. The droplets settle under gravity, grow by condensation, and nucleate during a specified time interval.

The droplet dynamics are described by a coupled momentum, heat, and mass-transfer model for an isolated water droplet in air \citep{Denner2026}, given as
\begin{align}
\dot{r} &= \frac{\mathrm{Sh}\,D_v M_w}{2r\rho_l R_g} \left[ \frac{p_{\mathrm{sat}}(T_\infty)}{T_\infty} - \frac{p_{\mathrm{sat}}(T)}{T} \right],
\\
\dot{w} &= -\frac{3\rho_aC_D\lvert w\rvert w}{8\rho_lr} - g\left(1-\frac{\rho_a}{\rho_l}\right),
\\
\dot{T} &= \frac{3}{r c_l} \left[ \frac{\mathrm{Nu}\,k_a(T_\infty-T)}{2r\rho_l} + L(T)\dot{r} \right].
\label{eq:applications-droplets}
\end{align}
The ambient air is assumed to be saturated, quiescent, and at a uniform temperature of $T_\infty=313.15$~K. The Sherwood number $\text{Sh}$ and the Nusselt number $\text{Nu}$ are computed based on the correlation of \citet{Beard1971} and the drag coefficient $C_\text{D}$ is based on the correlation of \citet{Clift1971}. More details about the model as well as all the used fluid properties of water and air can be found in the work of \citet{Denner2026}. In addition to the continuously evolving initial population, new droplets nucleate with $r_{\mathrm{nuc}}=5.0\times10^{-5}\,\mathrm{m}$, $w_\text{nuc}=0$, and $T_\text{nuc}=298.15\,\mathrm{K}$ during the time interval  $0.20 \, \text{s} \leq t < 0.40 \, \text{s}$. This nucleation contributes to a raw moment $M_{a,b,c}$ according to
\begin{equation}
\left. \frac{\mathrm{d}M_{a,b,c}}{\mathrm{d}t} \right|_{\mathrm{bnd}} = J(t) r_{\mathrm{nuc}}^a w_{\mathrm{nuc}}^b T_{\mathrm{nuc}}^c,
\label{eq:applications-droplets-nucleation}
\end{equation}
with the nucleation rate being
\begin{equation}
J(t)=
\begin{cases}
5 \times 10^4 \, \text{s}^{-1}, & 0.20\leq t<0.40~\mathrm{s},\\
0,   & \text{otherwise}.
\end{cases}
\label{eq:applications-droplets-nucleation-rate}
\end{equation}
The nucleation rate is chosen such that the initial droplet population doubles over the nucleation interval. The droplet radius is initialized with $r\sim\Gamma(16,18.75 \times 10^{-6} \, \mathrm{m})$, while the velocity and temperature are initialized with $w\sim\mathcal{N}(-2,1)\ \mathrm{m/s}$ and $T\sim\mathcal{N}(293.15,3)\ \mathrm{K}$, respectively. Hence, since the droplets are initially colder than their surrounding, the vapor concentration at the droplet surface is initially lower than in the ambient air and $\dot r>0$. QMOMKit uses the adaptive CQMOM implementation with a maximum of three nodes for each coordinate and applies the conditioning order $(r,T,w)$. 

Figure~\ref{fig:applications-droplets} compares the total number density, the characteristic radius $M_{3,0,0}/M_{2,0,0}$, the mean vertical velocity $M_{0,1,0}/M_{0,0,0}$, the mean temperature $M_{0,0,1}/M_{0,0,0}$, the mean product of radius and temperature $M_{1,0,1}/M_{0,0,0}$, as well as the radius variance
\begin{equation}
\sigma_r^2 = \frac{M_{2,0,0}}{M_{0,0,0}} - \left( \frac{M_{1,0,0}}{M_{0,0,0}}\right)^2.
\end{equation}
The imposed nucleation produces the expected linear doubling of the population between $0.2$ and $0.4$~s. At the same time, nucleation reduces the mean radius and broadens the radius distribution, resulting in a sharply increasing radius variance. The newly introduced droplets have zero vertical velocity and, consequently, reduce the mean velocity of the population during the injection interval. The entry temperature of the nucleated droplets hardly modifies the mean droplet temperature, whereas the mean product of radius and temperature decreases markedly, because the nucleated droplets are much smaller than the existing droplets. In general, the solution provided by QMOMKit closely follows the Monte Carlo reference for all six reported quantities, including the rapid changes during droplet nucleation.

\subsection{Spherical-particle breakup with diameter-dependent recoil}
\label{sec:applications-spherical-breakup-quadrature}

The first breakup example considers a population of spherical particles in a constant gas flow, described by the four-dimensional phase space $\boldsymbol{\xi}=(d,u,v,w)$, where $d$ is the particle diameter and $\mathbf{U}=(u,v,w)$ is the particle velocity. The initial population has unit number density, with
$\ln d \sim \mathcal{N}\!\left(\ln d_\text{ref},0.12^2\right)$ and 
$u,v,w \sim \mathcal{N}(0,0.5^2)$,
where $d_\text{ref} = 10^{-4}\,\mathrm{m}$. QMOMKit is configured with two quadrature nodes for each phase-space dimension and conditioning order $(d,u,v,w)$.

Breakup occurs at the constant rate $\lambda = 10\,\mathrm{s^{-1}}$, with each parent particle producing two equal-volume daughter particles, with daughter diameter $d_\pm=d/\sqrt[3]{2}$. The two daughters receive velocities subject to equal and opposite velocity increments, $\mathbf{U}_\pm = \mathbf{U} \pm \mathbf{R}$, where the components of the recoil vector $\mathbf{R}$ are independent Gaussian variables with zero mean and standard deviation $\sigma_R(d) = \sigma_0\, d / d_\text{ref}$ with $\sigma_0 = 0.25\,\mathrm{m/s}$. Because each daughter has half the parent volume, each individual breakup event preserves both the total particle volume and the volume-weighted momentum of the daughter pair. The size-dependence of the recoil amplitude, however, couples particle size and velocity in the breakup source and, hence, the source term cannot be evaluated solely using the finite set of moments advanced by CQMOM. Between breakup events, the particle diameter remains constant and the velocity evolves under Schiller-Naumann drag \citep{Schiller1933} towards the prescribed gas velocity $\mathbf{U}_\text{g} = (2,1.5,0.8)\,\mathrm{m/s}$. 

For a parent state $(d,u,v,w)$, the contribution of one breakup event to a raw moment is given by the combined contribution of the two daughters minus that of the parent. Multiplying this expected change by the breakup rate gives
\begin{equation}
\mathcal{S}_{p,q,r,s} = \lambda \left[\mathbb{E} \left( d_+^p u_+^q v_+^r w_+^s + d_-^p u_-^q v_-^r w_-^s \right) - d^p u^q v^r w^s \right],
\label{eq:applications-spherical-breakup-4d-local-source}
\end{equation}
where the expectation accounts for the random recoil. For a centered Gaussian variable $R$ with standard deviation $\sigma$, the required velocity moments can be evaluated analytically as
\begin{equation}
G_k(a,\sigma) = \mathbb{E}\!\left[(a+R)^k\right] = \sum_{\substack{\ell=0\\ \ell\ {\rm even}}}^{k} \binom{k}{\ell} a^{k-\ell} (\ell-1)!!\,\sigma^\ell, \qquad (-1)!!=1 .
\label{eq:applications-spherical-breakup-4d-gaussian-moment}
\end{equation}
Because the three recoil components are independent and centered, the expected daughter contribution separates into the three velocity directions, with Eq.~\eqref{eq:applications-spherical-breakup-4d-local-source} becoming
\begin{equation}
\mathcal{S}_{p,q,r,s} = \lambda \left[2\kappa^p d^p G_q\!\left(u,\sigma_R(d)\right) G_r\!\left(v,\sigma_R(d)\right) G_s\!\left(w,\sigma_R(d)\right) - d^p u^q v^r w^s \right],
\label{eq:applications-spherical-breakup-4d-local-source-gaussian}
\end{equation}
with $\kappa = d_{\pm}/d = 1/\sqrt[3]{2}$.
If $W_i$ denotes a terminal quadrature weight and $(d_i,u_i,v_i,w_i)$ are the corresponding phase-space abscissas, the breakup contribution to the moment rate is then
\begin{equation}
\left. \frac{\mathrm{d}M_{p,q,r,s}}{\mathrm{d}t} \right|_{\mathrm{br}} = \lambda \sum_i W_i \left[2\kappa^p d_i^p G_q\!\left(u_i,\sigma_R(d_i)\right) G_r\!\left(v_i,\sigma_R(d_i)\right) G_s\!\left(w_i,\sigma_R(d_i)\right) - d_i^p u_i^q v_i^r w_i^s \right].
\label{eq:applications-spherical-breakup-4d-source}
\end{equation}
Rather than expressing the breakup source solely in terms of the evolved moments, the parent-to-daughter transformation is evaluated directly on the reconstructed terminal quadrature.

The reason such a quadrature-based formulation of the breakup source term is useful becomes clear by considering one particular evolved moment. For the evolved moment $M_{1,3,0,0}$, the cubic velocity contribution is
\begin{equation}
G_3(u,\sigma_R) = u^3+3u\sigma_R^2 .
\end{equation}
Since the recoil variance scales with $d^2$,
$\sigma_R^2=\sigma_0^2d^2/d_{\mathrm{ref}}^2$, the source of the evolved
moment $M_{1,3,0,0}$ becomes
\begin{equation}
\left. \frac{\mathrm{d}M_{1,3,0,0}}{\mathrm{d}t} \right|_{\mathrm{br}} = \lambda \left[(2\kappa-1)M_{1,3,0,0} + 6 \kappa \frac{\sigma_0^2}{d_{\mathrm{ref}}^2} M_{3,1,0,0} \right].
\label{eq:applications-spherical-breakup-4d-untracked}
\end{equation}
Here the source of the tracked moment $M_{1,3,0,0}$ depends on $M_{3,1,0,0}$, which is not itself part of the evolved moment set using CQMOM with 2 nodes for each dimension. Evaluating the transformation on the terminal quadrature therefore provides a convenient route to obtain the required information. 

\begin{figure}[t]
\centering
\includegraphics[width=\linewidth]
{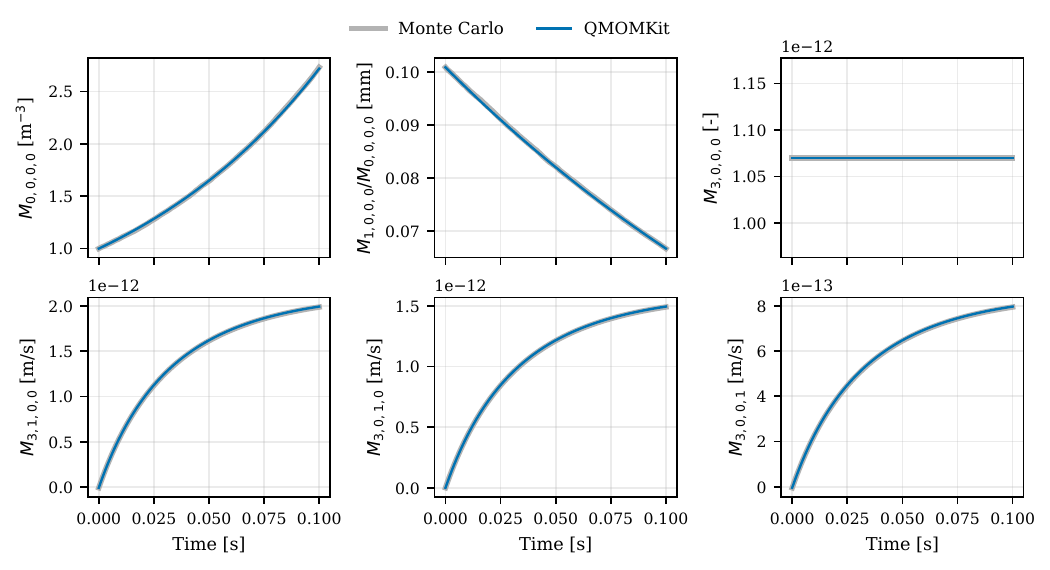}
\caption{Evolution of selected moments of a population of spherical particles subject to drag and constant-rate binary breakup in the four-dimensional phase space $(d,u,v,w)$, comparing fixed two-node CQMOM with the Monte Carlo reference. The top row shows the number density $M_{0,0,0,0}$, the mean diameter $M_{1,0,0,0}/M_{0,0,0,0}$, and the volume moment $M_{3,0,0,0}$. The lower row shows the volume-weighted velocity moments $M_{3,1,0,0}$, $M_{3,0,1,0}$, and $M_{3,0,0,1}$, which are not evolved by CQMOM but are evaluated from the quadrature. The considered equal-volume breakup conserves $M_{3,0,0,0}$, while drag accelerates the particles towards the imposed gas velocity.}
\label{fig:applications-spherical-breakup-quadrature}
\end{figure}

Figure~\ref{fig:applications-spherical-breakup-quadrature} compares the QMOMKit results with the corresponding Monte Carlo solution. For constant-rate binary breakup, each particle is replaced by two daughters, so the number density grows as $e^{\lambda t}$ and reaches approximately $e$ over the simulated interval. At the same time, the mean diameter decreases as successive breakup events produce smaller particles. The third size moment $M_{3,0,0,0}$ remains constant, which is the expected consequence of equal-volume breakup. The lower row of Figure~\ref{fig:applications-spherical-breakup-quadrature} shows the volume-weighted velocity moments $M_{3,1,0,0}$, $M_{3,0,1,0}$, and $M_{3,0,0,1}$. These moments are not part of the evolved moment set and are instead evaluated from the reconstructed terminal quadrature. The considered equal-volume breakup with symmetric recoil leaves them unchanged by the breakup event itself, while drag causes them to increase as the particle velocities approach $\mathbf{U}_\text{g}$. Their close agreement with the direct Monte Carlo reference therefore tests more than the evolved moments alone, showing that the reconstructed terminal population also provides an accurate representation of quantities required by the nonlinear source term.

\subsection{Spherical-particle breakup with heat transfer}
\label{sec:applications-spherical-breakup-moments}

Building on the previous example, this second breakup example includes heat transfer and demonstrates an alternative way of evaluating a moment source. Its five-dimensional phase space is $\boldsymbol{\xi}=(V,T,u,v,w)$, where $V$ is the particle volume, $T$ is the particle temperature, and $\mathbf{U}=(u,v,w)$ is the particle velocity. The initial volume distribution is obtained by transforming the diameter distribution used in the 4D example through $V=\pi d^3/6$, so that both cases represent the same initial particle-size distribution. The temperature and velocity distributions are initialized with $T\sim\mathcal{N}(350, 50^2)$~K and $u,v,w\sim\mathcal{N}(0,0.5^2)$ m/s, respectively. CQMOM is configured with two quadrature nodes for each dimension and conditioning order $(V,T,u,v,w)$.

Breakup again occurs at the constant rate $\lambda=10\,\mathrm{s^{-1}}$. Each parent produces two identical daughters in $(V,T)$, with $V'=V/2$ and 
\begin{equation}
T' = T + \varepsilon \left(\frac{V'}{V_{\mathrm{ref}}}\right)^{1/3} (T_\text{g}-T) = T + \varepsilon \left(\frac{V}{2V_{\mathrm{ref}}}\right)^{1/3} (T_\text{g}-T),
\label{eq:applications-spherical-breakup-5d-daughter}
\end{equation}
with gas temperature $T_\text{g}=300 \, \text{K}$, $\varepsilon=10^{-2}$ and $V_\text{ref} = \pi d^3_\text{ref}/6$, where $d_\text{ref} = 10^{-4}\,\mathrm{m}$. The velocity components remain unchanged by the breakup.  The important feature is the factor $V^{1/3}$ in Eq.~\eqref{eq:applications-spherical-breakup-5d-daughter}: When this daughter temperature is inserted into a moment containing powers of $T$, fractional powers of $V$ appear. To describe the corresponding fractional-order moments, we use the generalized-moment notation
\begin{equation}
\mathcal{M}_{p,q,r,s,t} = \int V^p T^q u^r v^s w^t n\, \mathrm{d}V\,\mathrm{d}T\,\mathrm{d}u\,\mathrm{d}v\,\mathrm{d}w , \qquad p\in\mathbb{R}_{\geq 0},
\label{eq:applications-spherical-breakup-5d-generalized-moment}
\end{equation}
which differs from the usual moment definition only in allowing the volume exponent $p$ to be non-integer. If $p$ is an integer, $\mathcal{M}$ is simply the corresponding moment $M$.  
Between breakup events, the particle volume remains constant, the velocity evolves under the same Schiller-Naumann drag law as in the previous example, and the temperature relaxes towards the gas temperature according to
\begin{equation}
\dot T = \frac{6k_\text{g}\text{Nu}} {\rho c_p d^2} \left(T_\text{g}-T\right),
\label{eq:applications-spherical-breakup-5d-thermal}
\end{equation}
where the Nusselt number $\text{Nu}$ is defined by the correlation of \citet{Ranz1952}.

The general breakup source follows by substituting Eq.~\eqref{eq:applications-spherical-breakup-5d-daughter} into the daughter moment and expanding the power of $T'$. The contribution of one daughter before accounting for the factor $V'=V/2$ is
\begin{equation}
\mathcal{D}_{p,q,r,s,t} = \sum_{\ell=0}^{q} \binom{q}{\ell} c^\ell \sum_{j=0}^{\ell} \binom{\ell}{j} T_\text{g}^{\ell-j}(-1)^j \mathcal{M}_{p+\ell/3,\,q-\ell+j,\,r,\,s,\,t},
\label{eq:applications-spherical-breakup-5d-expanded-daughter}
\end{equation}
with
\begin{equation}
c = \frac{\varepsilon}{\sqrt[3]{2V_{\mathrm{ref}}}}.
\end{equation}
Since one parent is replaced by two daughters of volume $V/2$, the resulting breakup source is
\begin{equation}
\left. \frac{\mathrm{d}M_{p,q,r,s,t}}{\mathrm{d}t} \right|_{\mathrm{br}} = \lambda \left(2^{1-p}\mathcal{D}_{p,q,r,s,t} - M_{p,q,r,s,t} \right).
\label{eq:applications-spherical-breakup-5d-source}
\end{equation}
The important point in Eq.~\eqref{eq:applications-spherical-breakup-5d-expanded-daughter}
is not the algebraic expansion itself, but the shifted volume order $p+\ell/3$. The closure mechanism is particularly transparent for the tracked moment $M_{1,1,0,0,0}$. Applying the daughter map gives
\begin{equation}
\mathcal{D}_{1,1,0,0,0} = M_{1,1,0,0,0} + c \left(T_\text{g} \mathcal{M}_{4/3,0,0,0,0} - \mathcal{M}_{4/3,1,0,0,0} \right),
\label{eq:applications-spherical-breakup-5d-temperature-daughter}
\end{equation}
and, because $2^{1-p}=1$ for $p=1$, the original $M_{1,1,0,0,0}$ contribution cancels when the parent is removed. The breakup source therefore reduces to
\begin{equation}
\left. \frac{\mathrm{d}M_{1,1,0,0,0}}{\mathrm{d}t} \right|_{\mathrm{br}} = \lambda c \left(T_g\mathcal{M}_{4/3,0,0,0,0} - \mathcal{M}_{4/3,1,0,0,0} \right).
\label{eq:applications-spherical-breakup-5d-temperature-source}
\end{equation}
This provides a direct example of source closure: CQMOM advances $M_{1,1,0,0,0}$ as part of the evolved moment set, but evaluating its breakup rate additionally requires $\mathcal{M}_{4/3,0,0,0,0}$ and $\mathcal{M}_{4/3,1,0,0,0}$. These fractional moments are calculated directly from the quadrature. The source can therefore remain written in moment form and, unlike the example presented in Section \ref{sec:applications-spherical-breakup-quadrature}, the calculation of the source term does not need to directly traverse the terminal quadrature nodes.

\begin{figure}[t]
\centering
\includegraphics[width=\linewidth]
{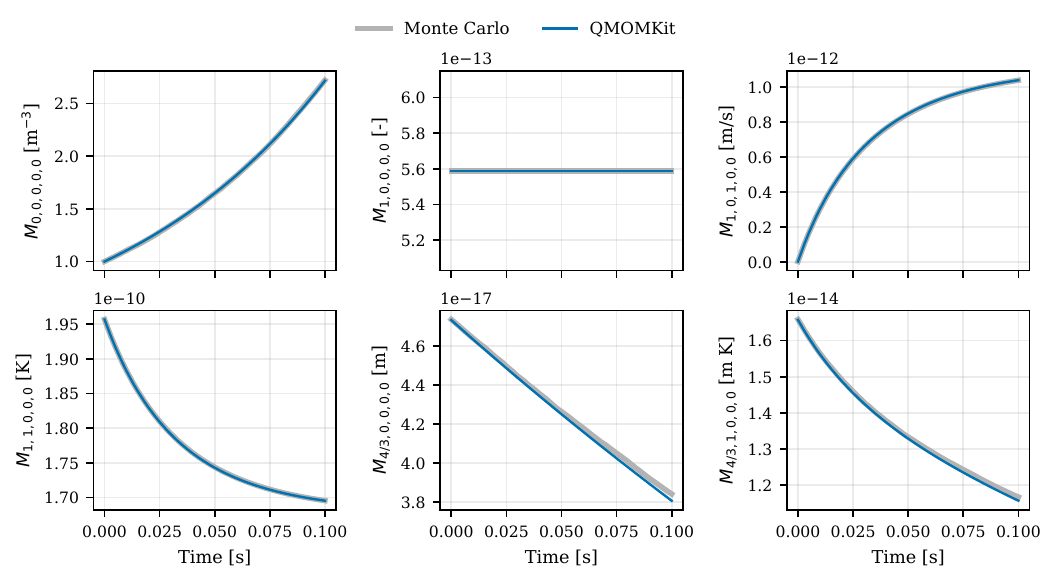}
\caption{Evolution of selected moments of a population of spherical particles subject to drag, heat transfer, and  constant-rate binary breakup in the five-dimensional phase space $(V,T,u,v,w)$, comparing fixed two-node CQMOM with the Monte Carlo reference. The top row shows the number density $M_{0,0,0,0,0}$, the volume moment $M_{1,0,0,0,0}$, and the volume-weighted velocity moment $M_{1,0,1,0,0}$. The lower row illustrates the fractional-moment source closure given in Eq.~\eqref{eq:applications-spherical-breakup-5d-temperature-source}: the evolved moment $M_{1,1,0,0,0}$ is shown together with the fractional moments $\mathcal{M}_{4/3,0,0,0,0}$ and $\mathcal{M}_{4/3,1,0,0,0}$ that are required to evaluate its breakup source. The small discrepancies of the fractional moments relative to the Monte Carlo reference arise because these moments are not matched directly by CQMOM, but are evaluated from the reconstructed quadrature and are therefore represented only approximately.}
\label{fig:applications-spherical-breakup-moments}
\end{figure}

Figure~\ref{fig:applications-spherical-breakup-moments} compares the QMOMKit results with the corresponding Monte Carlo solution. The top row first shows quantities whose behavior follows directly from the prescribed physics. The number density approaches $e^{\lambda t}=e$, while $M_{1,0,0,0,0}$ remains constant because it is proportional to the total particle volume. The moment $M_{1,0,1,0,0}$, which represents a volume-weighted velocity component, increases as drag accelerates the particle population towards the gas velocity. The lower row of Figure~\ref{fig:applications-spherical-breakup-moments} is chosen specifically to expose the fractional-moment closure in Eq.~\eqref{eq:applications-spherical-breakup-5d-temperature-source}. The left panel shows the tracked moment $M_{1,1,0,0,0}$ whose breakup source is being evaluated, while the center and right panels show the two fractional moments required by that source. The decrease of $M_{1,1,0,0,0}$ reflects
cooling of the particles towards the gas temperature together with the temperature change imposed during breakup. The fractional moment $\mathcal{M}_{4/3,0,0,0,0}$ decreases because a $4/3$ power of volume is not
conserved when one parent is replaced by two half-volume daughters, whereas $\mathcal{M}_{4/3,1,0,0,0}$ combines this size redistribution with the evolving particle temperature. The QMOMKit results closely follow the Monte Carlo reference for both fractional moments and for the tracked moment whose source depends on them.

\section{Coupling QMOMKit to a host application}
\label{sec:coupling}

Since QMOMKit is configured using CMake and does not have any external dependencies, it is straightforward to couple the library to existing simulation frameworks that use finite-volume or finite-element methods to solve PDEs. We demonstrate this by coupling QMOMKit with OpenFOAM v14 \cite{OpenFOAMv14}, an open-source software for computational fluid dynamics (CFD) based on the finite-volume method, and with deal.II v9.8 \cite{Arndt2026}, an open-source library to solve PDEs using finite-element methods. In the present work, we use the discontinuous Galerkin (DG) capabilities of deal.II. The example code for these couplings are available in a separate repository\footnote{https://github.com/polycfd/QMOMKit-Couplings}, as referenced in the QMOMKit repository.

\begin{figure}[t]
    \centering
    \includegraphics[width=0.5\linewidth]{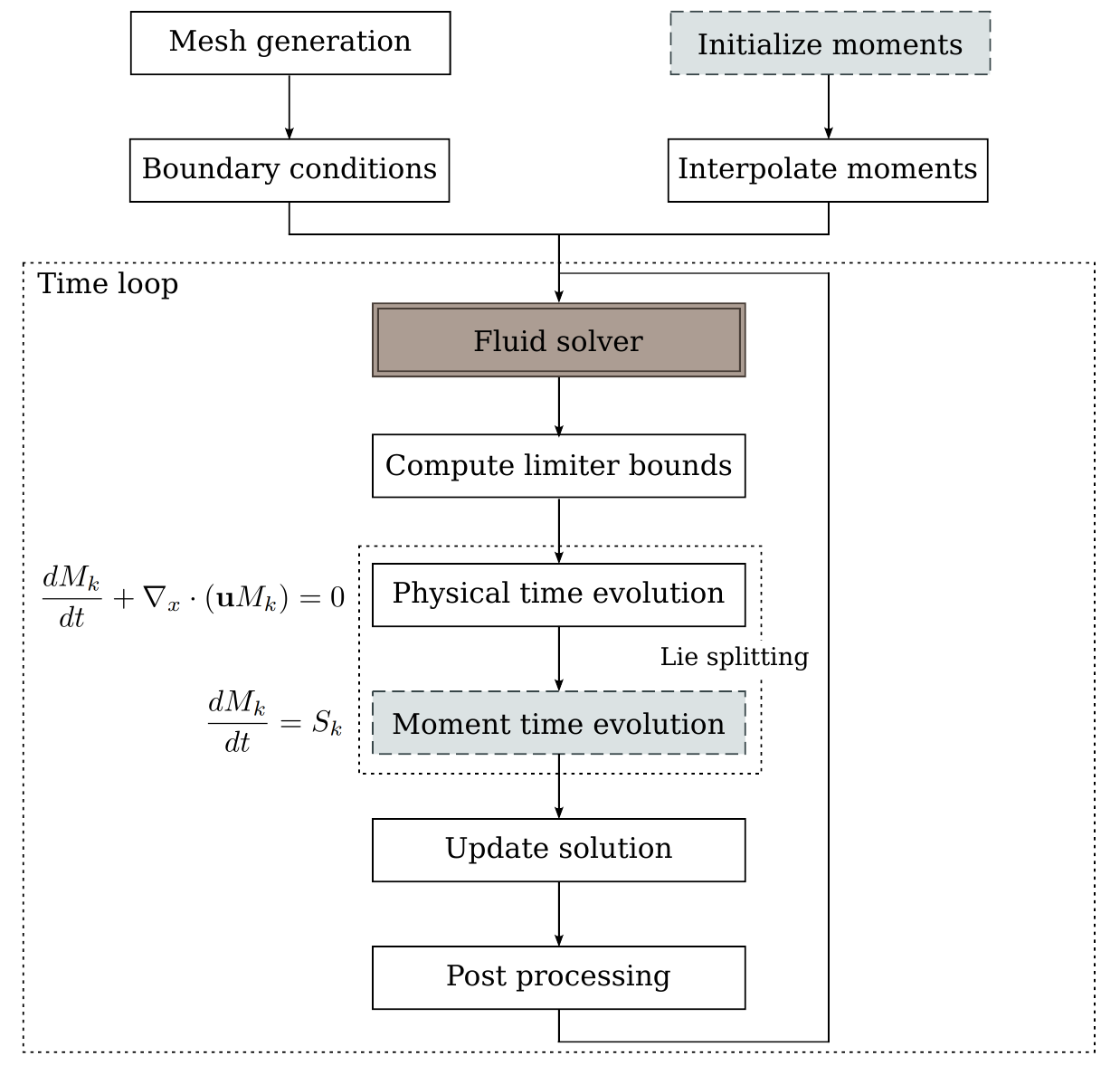}
    \caption{Flowchart for coupling of QMOMKit with the host applications of Section \ref{sec:coupling}, which here is either the finite-volume solver OpenFOAM or the finite-element DG formulation based on the deal.II library. The lightly shaded stages indicate the QMOMKit-dependent parts. The Lie splitting strategy employed for the time evolution have interchangeable physical and phase space stages.}
    \label{fig:coupling-chart}
\end{figure}

We consider a one-dimensional NDF $n(t,x,\xi)$ in physical and phase space, for which the governing PBE follows from Eq.~\eqref{eq:pbe} as
\begin{equation}
  \frac{\partial n}{\partial t} + \frac{\partial}{ \partial x} (n u)
  + \frac{\partial}{ \partial \xi} \left(n\,\dot{\xi}\right) = \mathcal{S},
\label{eq:problem_c_ndf}
\end{equation}
where $\xi$ represents a canonical phase-space coordinate, $u=1$ is a constant velocity, $\dot{\xi}$ is the velocity in phase space, and the source term is assumed to be $\mathcal{S}=0$. Multiplying Eq.~\eqref{eq:problem_c_ndf} by $\xi^k$ and integrating over
phase space gives
\begin{equation}
    \frac{\partial M_k}{\partial t} + \frac{\partial}{\partial x}\left(u M_k\right)
    - k \int_{-\infty}^{\infty} \xi^{k-1} \dot{\xi} n\,\mathrm{d}\xi = 0,
    \label{eq:problem_c_moment_general}
\end{equation}
where we assume no flux of $\xi$ through the boundary of phase space.
For a constant velocity $u=1$ and the constant growth term $\dot{\xi}=0.1$, the moment equations \eqref{eq:problem_c_moment_general} reduce to
\begin{equation}
    \frac{\partial M_k}{\partial t} + u \frac{\partial M_k}{\partial x} - k\, \dot{\xi} \, M_{k-1} = 0.
    \label{eq:problem_c_moments}
\end{equation}
In practice, $u$ would be obtained from the solution of a CFD problem, whereas here the goal is primarily to demonstrate the coupling between QMOMKit and a host code. 

The physical domain is a periodic domain of length 2, such that $x\in[-1,1]$.
We consider an initial NDF that is separable and Gaussian in both physical  and phase space,
\begin{equation}
    n(0,x,\xi)
    =
    g_x(x)\,
    g_\xi(\xi),
\end{equation}
with
\begin{equation}
    g_x(x)
    =
    \exp\left(
        -\frac{x^2}{2\sigma_x^2}
    \right),
    \qquad
    g_\xi(\xi)
    =
    \frac{1}{\sqrt{2\pi}\sigma}
    \exp\left[
        -\frac{(\xi-\mu)^2}{2\sigma^2}
    \right],
    \label{eq:problem_c_initial_ndf}
\end{equation}
where $\sigma_x = 0.25$, $\sigma = 1$, and $\mu=10$.
Consequently, the physical-space Gaussian is translated at constant velocity, while the phase-space Gaussian retains its variance and its mean increases linearly with time. The analytical NDF is therefore
\begin{equation}
    n(t,x,\xi)
    =
    g_x(x-ut)\,
    \frac{1}{\sqrt{2\pi}\sigma}
    \exp\left[
        -\frac{\left(\xi-(\mu + \dot{\xi} t)\right)^2}{2\sigma^2}
    \right],
    \label{eq:problem_c_analytical_ndf}
\end{equation}
where $x-t$ is understood periodically on the physical domain
$x\in[-1,1]$. It follows that the analytical moments can be written as
\begin{equation}
    M_k(t,x)
    =
    g_x(x-ut)\,
    \mathcal{M}_k(\mu + \dot{\xi} t,\sigma),
    \label{eq:problem_c_analytical_moments}
\end{equation}
where $\mathcal{M}_k(\bar{\xi},\sigma)$ denotes the $k$th raw moment of a normal distribution with mean $\bar{\xi}$ and standard deviation $\sigma$. The analytical solution for all moments is thus:
\begin{equation}
    M_k(t,x)
    =
    g_x(x-ut)
    \sum_{j=0}^{\lfloor k/2 \rfloor}
    \frac{k!}{2^j j!(k-2j)!}
    \sigma^{2j}\left(\mu + \dot{\xi} t\right)^{k-2j}.
\end{equation}

We solve Eq.~\eqref{eq:problem_c_moments} using a Lie operator splitting, as illustrated in Figure \ref{fig:coupling-chart}. The solution is advanced from $t^n$ to $t^{n+1}=t^n+\Delta t$ in two successive steps. First, the physical-space advection is solved,
\begin{equation}
    \frac{\partial M_k^{*}}{\partial t}
    +
    u \frac{\partial M_k^{*}}{\partial x}
    = 0,
    \qquad
    M_k^{*}(t^n)=M_k^n,
    \label{eq:host-coupling_physical-space-advection}
\end{equation}
over one time step $\Delta t$. The resulting intermediate solution is then used as the initial condition for the phase-space evolution with QMOMKit,
\begin{equation}
    \frac{\text{d} M_k}{\text{d} t}
    = k\, \dot{\xi} \, M_{k-1},
    \qquad
    M_k(t^n)=M_k^{*}(t^{n+1}),
\end{equation}
which is also advanced over $\Delta t$. We track the first 10 moments using a 5-node quadrature with QMOM. While the phase-space evolution is managed and solved by QMOMKit, the physical-space evolution is managed and solved by either OpenFOAM or deal.II. 
For OpenFOAM, a first-order upwind scheme is used for the spatial discretization of Eq.~\eqref{eq:host-coupling_physical-space-advection}, whereas the deal.II implementation uses first- and second-order discontinuous Lagrange elements (DG-$Q_p$, with polynomial degree $p$) combined with an upwind flux. The goal of this section is not to compare finite-volume and DG methods, but to highlight the capacity of QMOMKit to be readily coupled to both frameworks.

\begin{figure}[t]
    \centering
    \includegraphics[width=1\linewidth]{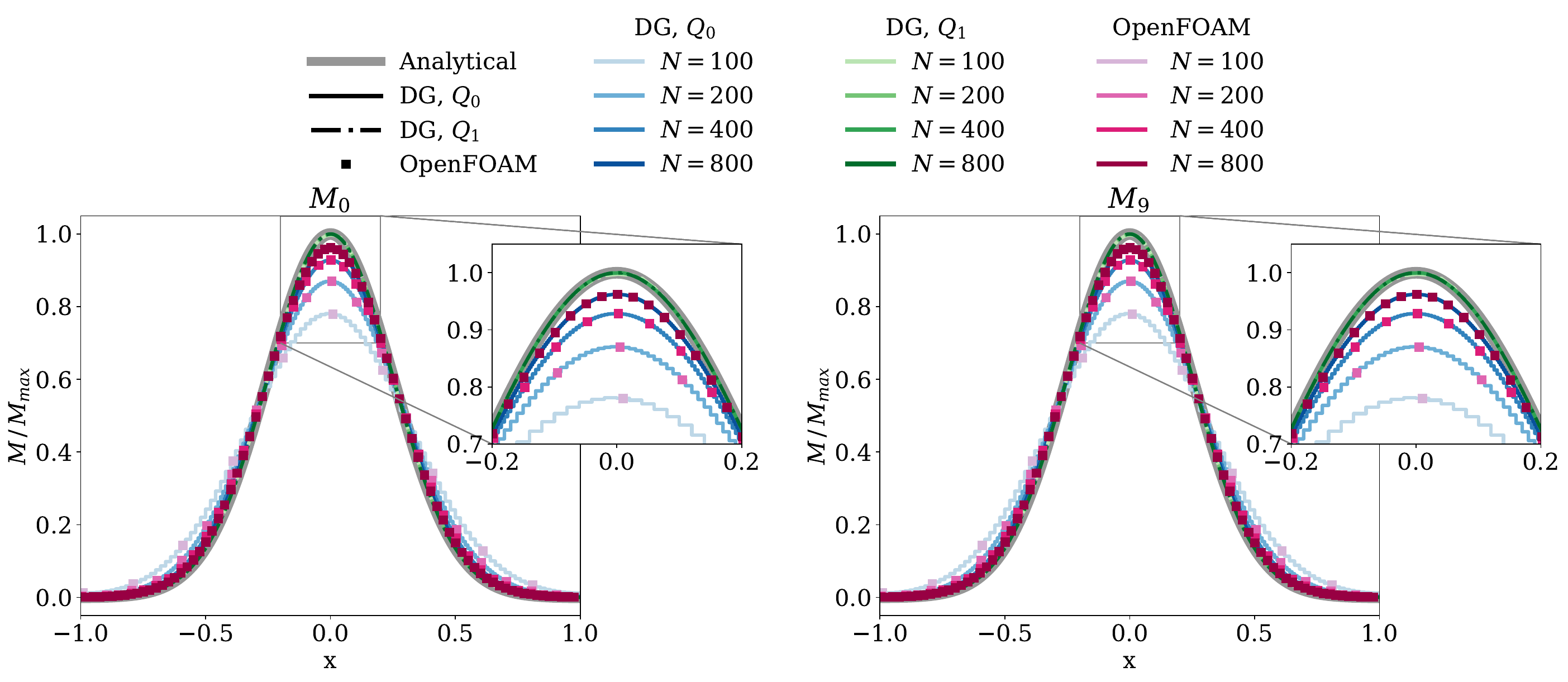}
    \caption{Representation of selected moments in physical space for the coupling of QMOMKit with two types of host applications: OpenFOAM and a deal.II-based DG solver. The moments are normalized by $M_\text{max}$, which is the maximum value of the analytical solution. Mesh convergence yields very similar results for the finite-volume discretization of OpenFOAM and the DG-$Q_0$ discretization for all mesh resolutions. The solution obtained with the DG-$Q_1$ discretization is in excellent agreement with the analytical solution, even for the coarsest mesh.}
    \label{fig:dg-coupling-mesh-conv}
\end{figure}

\begin{figure}[t]
    \centering
    \includegraphics[width=1\linewidth]{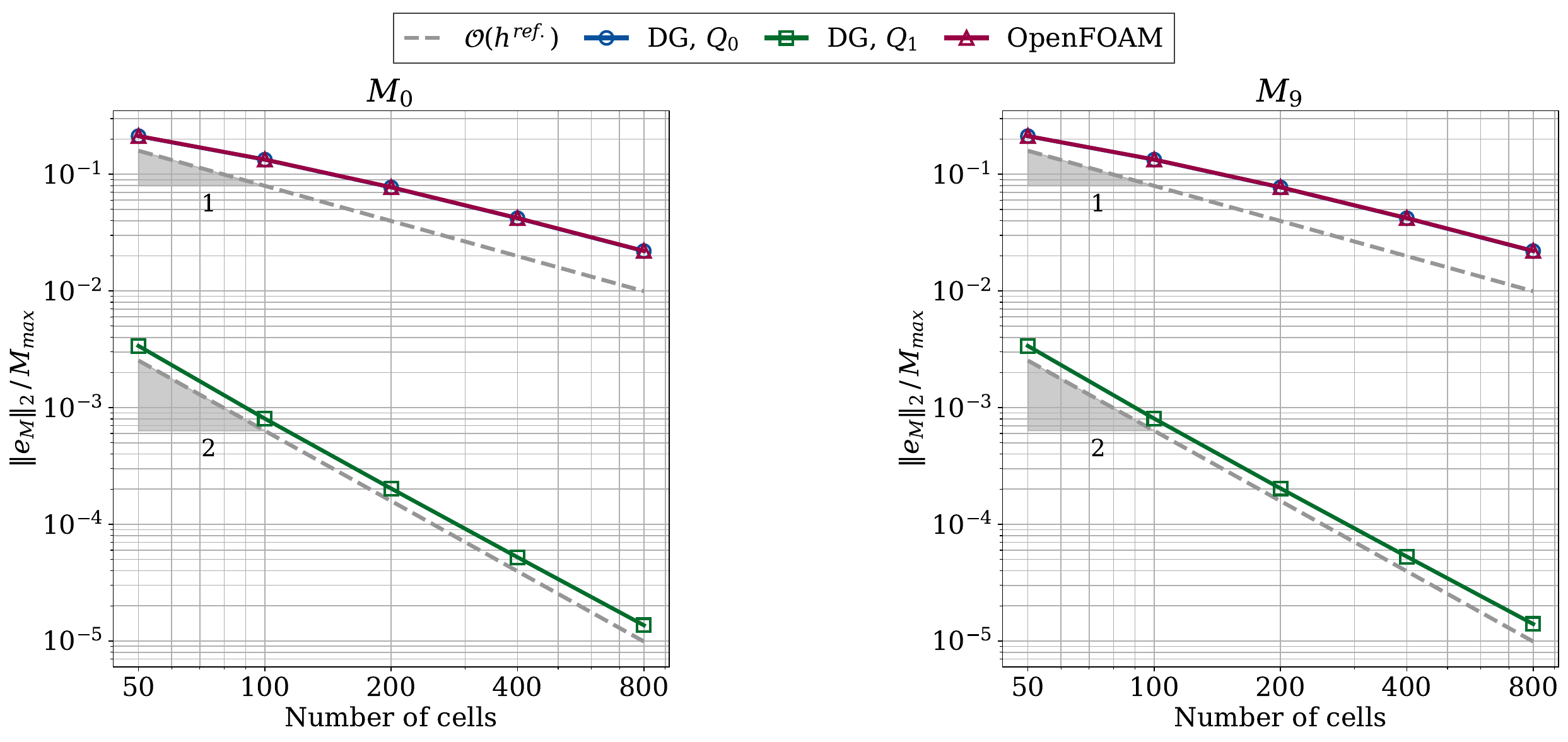}
    \caption{$\mathcal{L}^2$ norm of the error for selected moments normalized by $M_\text{max}$ (maximum value of the analytical solution). The solution errors obtained with the finite-volume discretization of OpenFOAM as well as the DG-$Q_0$  discretization of deal.II converge asymptotically with first order. In contrast, the DG-$Q_1$ discretization achieves second-order convergence.}
    \label{fig:dg-coupling-error}
\end{figure}

Figure \ref{fig:dg-coupling-mesh-conv} illustrates the results obtained after $2 \, \text{s}$, corresponding to the transport of the NDF over the full periodic domain. Quasi-identical results are obtained with the first-order upwind finite-volume and DG-$Q_0$ discretizations. The results obtained with the formally second-order accurate DG-$Q_1$ discretization exhibits a much better agreement with the analytical solution, even for the coarsest mesh considered.
Figure \ref{fig:dg-coupling-error} shows the evolution of the $\mathcal{L}^2$ norm of the error of the moments $M_0$ and $M_9$ as the mesh is progressively refined from $50$ to $800$ cells, while the CFL is kept constant at a value of $0.005$. Even though Lie splitting is applied, the nature of the problem allows DG-$Q_1$ elements to reach second-order convergence, whereas DG-$Q_0$ and OpenFOAM with an upwind discretization exhibit first-order convergence. 

\section{Conclusions}
\label{sec:conclusions}

In this work, we have presented QMOMKit, an open-source C++ library for quadrature-based moment methods. While the implemented QMOM and CQMOM formulations are well established, QMOMKit introduces a general numerical framework that removes several restrictions commonly encountered in application-specific implementations. In particular, the recursive formulation of CQMOM permits quadrature reconstruction in phase spaces of arbitrary dimension, whereby a common evolution framework is used for phase-space advection, moment-source evolution, and their operator-split combination. Together with configurable inversion algorithms, realizability checks, and robust treatment of degenerate or weakly populated quadratures, this provides a common computational framework for problems ranging from one-dimensional to high-dimensional conditional quadrature formulations.

QMOMKit separates the quadrature-based phase-space evolution from the surrounding physical model and spatial discretization. The host application supplies phase-space velocities, source terms, and boundary contributions through callbacks, whereas QMOMKit controls the quadrature reconstruction and evaluation of the callbacks during time integration. The resulting software implementation allows the same numerical algorithms to be used for spatially homogeneous population models or locally within finite-volume, finite-element, or other host solvers, without embedding application-specific assumptions in the library. The numerical implementation is supported by complementary unit, regression, and verification tests and by application examples spanning increasingly complex phase spaces and physical processes.

The present implementation also provides a foundation for future extensions of QMOMKit. In particular, the modular reconstruction framework can accommodate additional quadrature formulations, such as generalized QMOM (GQMOM) \citep{Fox2023a} and hyperbolic or conditional hyperbolic quadrature methods (HyQMOM or CHyQMOM) \citep{Fox2018,Fox2022}, while preserving the same application-facing interfaces for phase-space evolution. Preliminary work on incorporating GQMOM and HyQMOM is already under way. Additional inversion algorithms, procedures to preserve realizability, and time-integration schemes can likewise be incorporated within the existing framework. QMOMKit therefore provides not only a reusable implementation of the methods considered here, but also a common numerical and verification framework for the continued development of quadrature-based moment methods.

\section*{Acknowledgements}
We acknowledge the support of the Natural Sciences and Engineering Research Council of Canada (NSERC), funding reference numbers RGPIN-2020-04510 and RGPIN-2024-04805, the Canada Research Chair CRC-2022-00340, and the Deutsche Forschungsgemeinschaft (DFG, German Research Foundation), grant number 458610925. We also wish to thank Mohamed Amine Bouguezzoul for fruitful discussions.

\section*{Declaration of generative AI and AI-assisted technologies}
During the preparation of this work the authors used ChatGPT (OpenAI) and Claude (Anthropic) in order to improve the readability of this manuscript. After using this tool, the authors reviewed and edited the content as needed and take full responsibility for the content of the published article. 

\appendix
\section{Monte Carlo companion code}
\label{sec:monte-carlo}

QMOMKit is distributed with an optional Monte Carlo companion package for constructing numerical reference solutions, which is built separately and links against QMOMKit. The Monte Carlo companion initializes a finite ensemble of phase-space samples from prescribed distributions, such as univariate distributions, mixtures, and independent products of univariate distributions, and evolves each sample according to application-defined phase-space dynamics \citep{Irizarry2008}. Moments are then evaluated directly from the ensemble.

Let $n^{\mathrm{MC}}(t,\boldsymbol{\xi})$ denote the Monte Carlo representation of the NDF. For $N_{\mathrm{s}}$ samples with phase-space coordinates $\boldsymbol{\xi}_j(t)$ and equal representative weight $n_0/N_{\mathrm{s}}$,
\begin{equation}
n^{\mathrm{MC}}(t,\boldsymbol{\xi}) = \frac{n_0}{N_{\mathrm{s}}} \sum_{j=1}^{N_{\mathrm{s}}} \delta\!\left(\boldsymbol{\xi}-\boldsymbol{\xi}_j(t)\right).
\label{eq:mc-empirical-distribution}
\end{equation}
The raw moments are obtained directly from the ensemble as
\begin{equation}
\widehat{M}_{\boldsymbol{\alpha}}^{\mathrm{MC}} = \frac{n_0}{N_{\mathrm{s}}} \sum_{j=1}^{N_{\mathrm{s}}} \prod_{d=1}^{D}\xi_{j,d}^{\alpha_d}. \label{eq:mc-raw-moment}
\end{equation}
In particular, $\widehat{M}_{\boldsymbol{0}}^{\mathrm{MC}}=n_0$. Because the ensemble is finite, the moment estimates contain a sampling error, which decreases with $1/\sqrt{N_{\mathrm{s}}}$.

With $\dot{\boldsymbol{\xi}}$ supplied by the application through the same callback as for QMOMKit, each sample evolves independently according to
\begin{equation}
\frac{\mathrm{d}\boldsymbol{\xi}_j}{\mathrm{d}t} = \dot{\boldsymbol{\xi}} \left(t,\boldsymbol{\xi}_j\right), \qquad j=1,\ldots,N_{\mathrm{s}}.
\label{eq:mc-sample-evolution}
\end{equation}
The function \texttt{evolve\_ensemble()} advances these trajectories using the embedded RK23 scheme \citep{Bogacki1989}, with an adaptive step size for each sample. Unlike QMOMKit, which evolves quadrature abscissas, the Monte Carlo companion advances every sample explicitly.

Processes that change the population are represented by changes to the population ensemble. Population additions are supported directly through \texttt{inject\_population()}, which draws a Poisson-distributed number of new samples from an application-supplied distribution. The new samples receive the existing representative weight, so that the requested increase of the number density is reproduced in expectation. Other processes, such as breakage, are defined by the application by sampling the corresponding population events \citep{Lee2015}. In the breakup examples of Sections~\ref{sec:applications-spherical-breakup-quadrature} and \ref{sec:applications-spherical-breakup-moments}, the time until breakup is drawn from an exponential distribution for each particle, and the daughters keep the representative weight, so that the ensemble grows. These operations provide the Monte Carlo counterpart to moment-source and boundary contributions in the QMOMKit calculations. Sampling, injection, and breakup use \texttt{std::mt19937\_64} generators with fixed seeds, so that calculations can be reproduced with the same compiler and standard library. 

In this work, the Monte Carlo calculations use the same physical models, time step, operator splitting, and RK23 tolerances as the QMOMKit simulations. Moreover, the QMOMKit simulations are initialized with the moments of the Monte Carlo ensemble, so that both calculations start from the same population. The comparisons in Section~\ref{sec:applications} therefore measure the error of the employed quadrature-based moment methods. Without population sources, as in Sections~\ref{sec:applications-aerosol} and \ref{sec:applications-cavitation}, the Monte Carlo calculation is deterministic and simply evolves the population used to initialize the corresponding QMOMKit simulation. Randomly sampled events, such as breakup or nucleation, add a statistical error that decreases with $1/\sqrt{N_{\mathrm{s}}}$.


\end{document}